\documentclass[useAMS,usenatbib]{mn2e}
\usepackage{amssymb,amstext,amsmath,commath}
\usepackage{times,float,graphicx,paralist}
\usepackage{epsfig,epstopdf,url,rotating}
\DeclareGraphicsExtensions{.pdf,.ps,.eps,.png,.jpg,.mps} 
\thinspace
\begin{document}
\title[Radial Profiles of Spiral Galaxies]{The JCMT Nearby Galaxies Legacy Survey -- XI. -- Environmental Variations in the Atomic and Molecular Gas Radial Profiles of Nearby Spiral Galaxies}
\author[A. Mok et al.] {Angus Mok $^{1}$, C. D. Wilson$^{1}$, J. H. Knapen$^{2,3}$, J. R. S\'anchez-Gallego$^{4}$,
\newauthor E. Brinks$^{5}$, and E. Rosolowsky$^{6}$ \\
\\
$^{1}$ Department of Physics \& Astronomy, McMaster University, Hamilton, Ontario L8S 4M1, Canada \\
$^{2}$ Instituto de Astrof\'isica de Canarias, E-38205 La Laguna, Tenerife, Spain \\
$^{3}$ Departamento de Astrof\'\i sica, Universidad de La Laguna, E-38200 La Laguna, Tenerife, Spain \\
$^{4}$ Department of Physics and Astronomy, University of Kentucky, 40506-0055, Lexington, Kentucky, United States \\
$^{5}$ Centre for Astrophysics Research, University of Hertfordshire, AL10 9AB, Hatfield, Hertfordshire, United Kingdom \\
$^{6}$ Department of Physics, University of Alberta, Edmonton, Ontario T6G 2R3, Canada \\
}
%
%
\def\etal{{ et al.\thinspace}}
\def\gtrsim{\mathrel{\raise0.35ex\hbox{$\scriptstyle >$}\kern-0.6em\lower0.40ex\hbox{{$\scriptstyle \sim$}}}}
\def\lesssim{\mathrel{\raise0.35ex\hbox{$\scriptstyle <$}\kern-0.6em\lower0.40ex\hbox{{$\scriptstyle \sim$}}}}
\def\Msun{\hbox{$\rm\thinspace M_{\odot}$}}
\def\kmsmpc{{\,\rm km\,s^{-1}Mpc^{-1}}}
\def\kms{km~s$^{-1}$}
\def\hi{H\,{\sc i\,}}
\def\h2{H$_{2}$}
\def\ha{H$\alpha$}
\def\ks{Kolmogorov-Smirnov }
\def\co{CO $J=3-2$}
\def\aap{A\&A}
\def\aaps{A\&AS}
\def\araa{ARA\&A}
\def\apjl{ApJ}
\def\apj{ApJ}
\def\apjs{ApJS}
\def\aj{AJ}
\def\mnras{MNRAS}
\def\pasj{PASJ}
\def\pasp{PASP}
\def\nat{Nature}
%
%
\maketitle
\begin{abstract}
We present an analysis of the radial profiles of a sample of 43 \hi-flux selected spiral galaxies from the Nearby Galaxies Legacy Survey (NGLS) with resolved James Clerk Maxwell Telescope (JCMT) \co\ and/or Very Large Array (VLA) \hi\ maps. Comparing the Virgo and non-Virgo populations, we confirm that the \hi\ disks are truncated in the Virgo sample, even for these relatively \hi-rich galaxies. On the other hand, the \h2\ distribution is enhanced for Virgo galaxies near their centres, resulting in higher \h2\ to \hi\ ratios and steeper \h2\ and total gas radial profiles. This is likely due to the effects of moderate ram pressure stripping in the cluster environment, which would preferentially remove low density gas in the outskirts while enhancing higher density gas near the centre. Combined with \ha\ star formation rate data, we find that the star formation efficiency (SFR/\h2) is relatively constant with radius for both samples, but Virgo galaxies have a $\sim40\%$ lower star formation efficiency than non-Virgo galaxies. 
\end{abstract}
\begin{keywords}
galaxies: ISM -- galaxies: spiral -- ISM: molecules -- stars: formation
\end{keywords}
%
%
\section{Introduction}
\par
In the Universe, galaxies can be found in many different environments, from isolated galaxies, to groups of tens of galaxies, to clusters of thousands of galaxies. The star formation histories of these galaxies have been observed to vary greatly due to their environment. For example, overdense regions are increasingly dominated by quiescent galaxies \citep{Dressler80, Blanton09}. One of the major questions in the study of galaxy evolution is whether these effects are due to processes intrinsic to the galaxy, such as through the process of `mass quenching', or if they can be attributed to the local environment \citep{Peng10, Peng12}. The lower density group environment is also important in this analysis, as galaxies can begin to be quenched in these groups before falling into clusters \citep{McGee09}.
\par
Optical studies of galaxies at a range of redshifts have provided valuable insight into these environmental effects, but a full analysis requires the study of the fuel for ongoing and future star formation. Spiral galaxies are important to this analysis because they are sites of active star formation, with gaseous disks that are sensitive to the effects of their surroundings. Past studies of the \hi\ content of spiral galaxies in clusters have found that they are deficient in atomic gas \citep{Haynes86} and their gas distributions are significantly truncated \citep{Cayatte90}.
\par
More recent studies have shown that star formation is most closely linked to the molecular gas content in these galaxies \citep{Leroy08, Saintonge11b, Bigiel11}. Given the observed reduction in overall star formation inside cluster spirals \citep{Koopmann06}, the cluster environment should therefore have a strong effect on their molecular gas content. Processes such as ram pressure stripping \citep{Gunn72}, tidal harassment \citep{Moore96}, and strangulation \citep{Larson80} have been proposed to explain the environment's influence on the more diffuse \hi\ component. However, any potential environmental effects on the denser, more centrally located \h2\ gas is less clear.
\par
Initial observations by \citet{Kenney89} for 40 Virgo spirals using the Five College Radio Astronomy Observatory (FCRAO) found that they are not as \h2\ deficient as they are \hi\ deficient and suggested the survival of a large amount of molecular gas inside these galaxies. More recently, the Herschel Virgo Cluster Survey (HeViCS) created a sample of 12 galaxies with CO measurements and found that for disturbed galaxies, the \hi\ gas is more efficiently stripped than the \h2\ gas \citep{Pappalardo12}. This results in steeper \h2\ and total gas radial profiles for the more \hi-deficient galaxies in their sample. The Herschel Reference Survey (HRS), which is a K-band selected sample of nearby galaxies, used the \hi-deficiency parameter as a proxy for galaxy interactions and found a correlation between \hi-deficiency and the level of \h2\ gas deficiency and \h2\ gas disk size \citep{Boselli14b}. They suggested that Virgo spiral galaxies may be more \h2 deficient than unperturbed field galaxies.
\par
Here, we take a slightly different approach by using a gas-rich sample of spiral galaxies in three different environments (field, group, and the Virgo Cluster), which should be better proxies for any immediate environmental effects. To study the molecular gas properties of these galaxies, we use resolved maps from the Nearby Galaxies Legacy Survey (NGLS), an \hi-flux limited sample of 155 nearby ($D < 25$ Mpc) galaxies \citep{Wilson12}. In a previous paper, we select spiral galaxies from the NGLS sample to compare the field, group, and Virgo subsamples, focusing on their integrated properties \citep{Mok16}. Compared to non-Virgo galaxies, galaxies in the Virgo cluster have higher \h2\ gas masses and \h2\ to \hi\ ratios, perhaps due to environmental interactions that aid in the conversion of atomic to molecular gas. They also have lower specific star formation rates (${\rm sSFR} = {\rm SFR}/M_*$) and lower star formation efficiencies (${\rm SFR}/M_{\rm H_2}$), implying molecular gas depletion times ($t_{\rm gas}= M_{\rm H_2}/{\rm SFR}$) that are longer than non-Virgo galaxies. 
\par
In this paper, we will further investigate the environmental effects by studying the radial profiles of the NGLS galaxies and measure any differences between the Virgo and non-Virgo sample. We select 43 galaxies from the larger sample of NGLS spiral galaxies for which we have detected maps of their \h2 and/or \hi\ distributions, while excluding galaxies with inclinations of greater than 75 degrees. In \S~2, we present our observations and some of the general properties of our sample. In \S~3, we discuss the procedure used to create the radial profiles. We also present radial profiles for the \h2, \hi, and star formation rate surface densities, as well as combinations of these properties, such as the molecular gas depletion time. Then, we discuss some implications of our data and provide comparisons to other observational and theoretical results.
%
%
\section{Observations and Analysis}
\subsection{Data and Sample Selection}\label{sec-dataselection}
\par
We seek to create a large sample of spiral galaxies with resolved molecular and/or atomic gas data available. To the original Nearby Galaxies Legacy Survey (NGLS) sample \citep{Wilson12}, we add galaxies that match the NGLS survey criterion from two follow-up JCMT programs. The first is a program to complete the \hi-flux selected sample of galaxies in the Virgo Cluster (M09AC05) and the second is to map the galaxies from the Herschel Reference Survey (M14AC05). From this larger sample, we select only spiral galaxies using the HyperLeda database, excluding any elliptical or lenticular galaxies. This results in the 98 spiral galaxies used in the analysis from \citet{Mok16}. For all of the galaxies in the survey, the CO $J=3-2$ line was mapped out to at least $D_{25}/2$ using HARP on the James Clerk Maxwell Telescope (JCMT) \citep{Wilson12}. The $D_{25}$ value for each galaxy, which we take from the HyperLEDA survey, is defined as the length of the major axis of a galaxy at the B-band isophotal level of 25 mag/arcsec2 \citep{Paturel03}. In our analysis, we also use $R_{25}$, which is half of the $D_{25}$ value for each galaxy.
\par
To study the resolved properties of these galaxies, we select only galaxies for which we have detections in the \co\ maps from the JCMT and/or the \hi\ 21 cm line from the VLA. We exclude all galaxies with inclinations of greater than 75 degrees, as measured by the HyperLeda database \citep{Paturel03, Makarov14}, to remove any galaxies that are close to an edge-on configuration and not appropriate for our radial profile analysis. Our final resolved sample contains 43 galaxies.
\par
In total, there are 33 galaxies with \co\ detections from the JCMT. One of the main advantages to our molecular gas dataset is that the observations were all made using the same instrument, in the same CO transition, and with the same survey specifications. This homogeneous dataset provides a good basis for comparison between the different galaxies and different environments. The \co\ data reduction process is described in \citet{Wilson12} and will not be repeated here. A full analysis of the integrated properties of the spiral galaxies in this sample can be found in \citet{Mok16}.
\par
\begin{figure*}
	\centering
	AW701:\\
	\includegraphics[width=5.75cm]{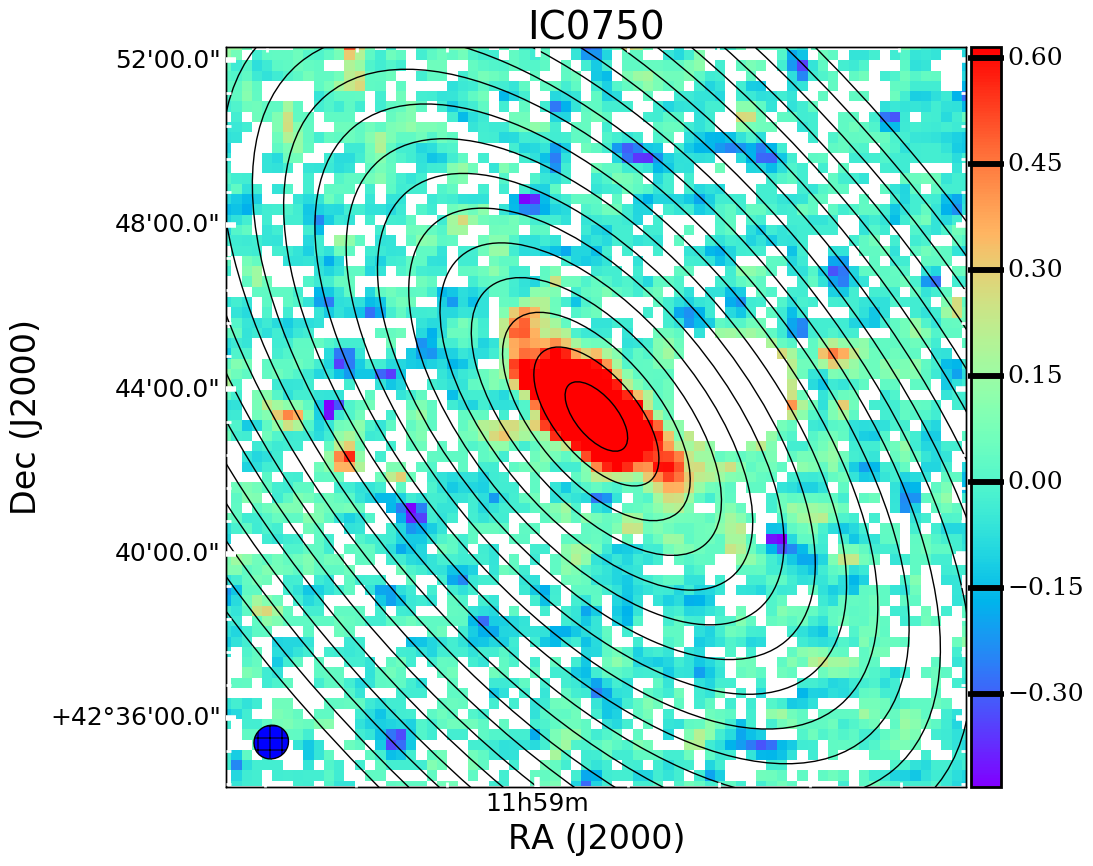}
	\includegraphics[width=5.75cm]{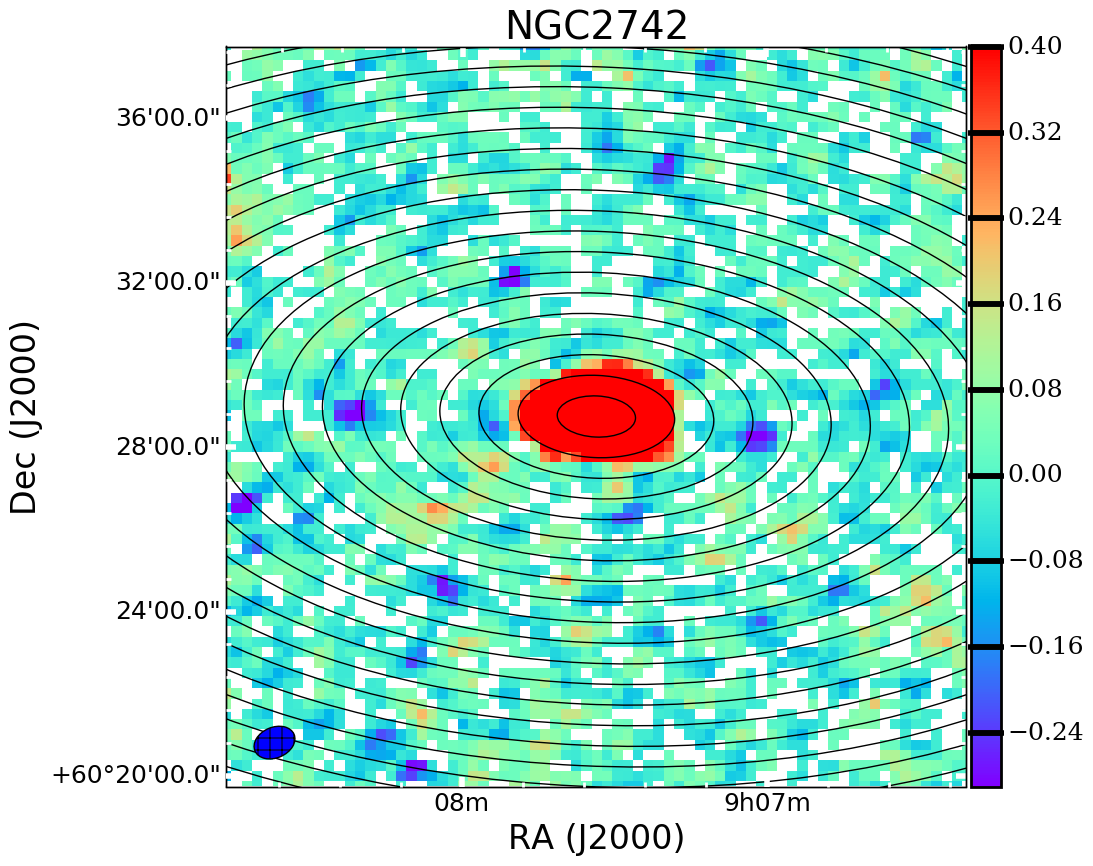}
	\includegraphics[width=5.75cm]{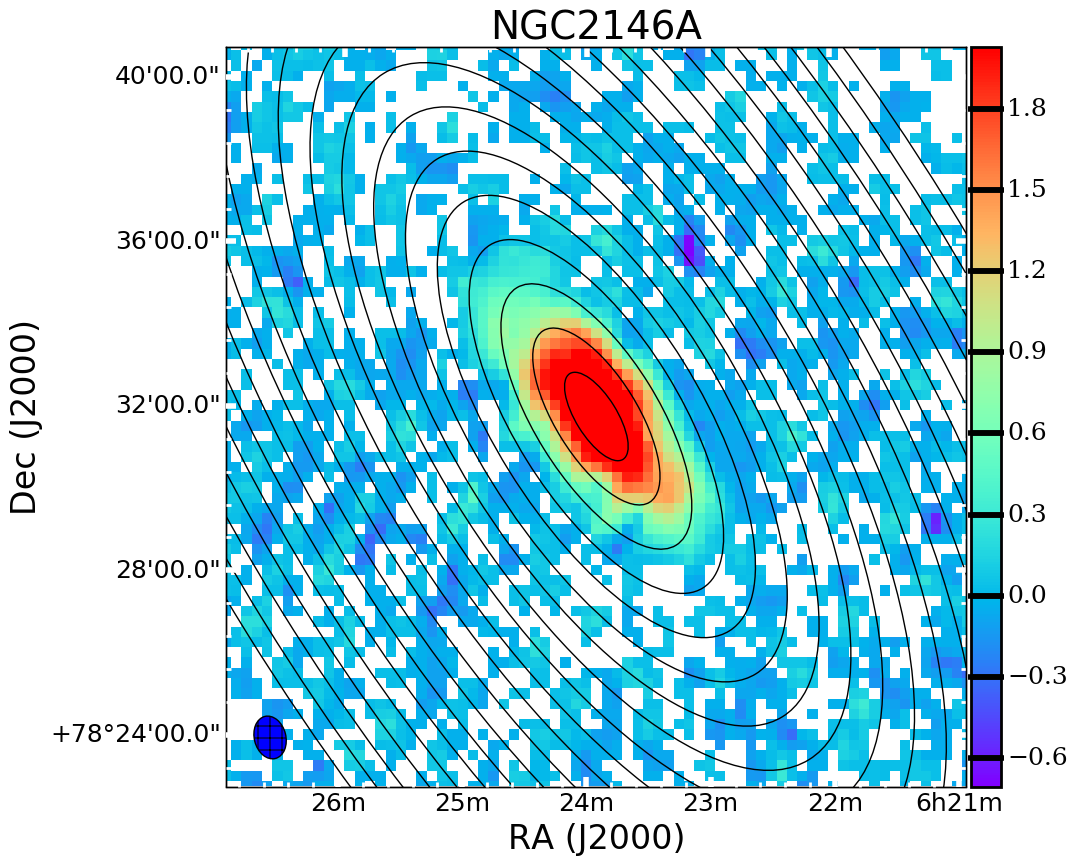}
	\\
	\includegraphics[width=5.75cm]{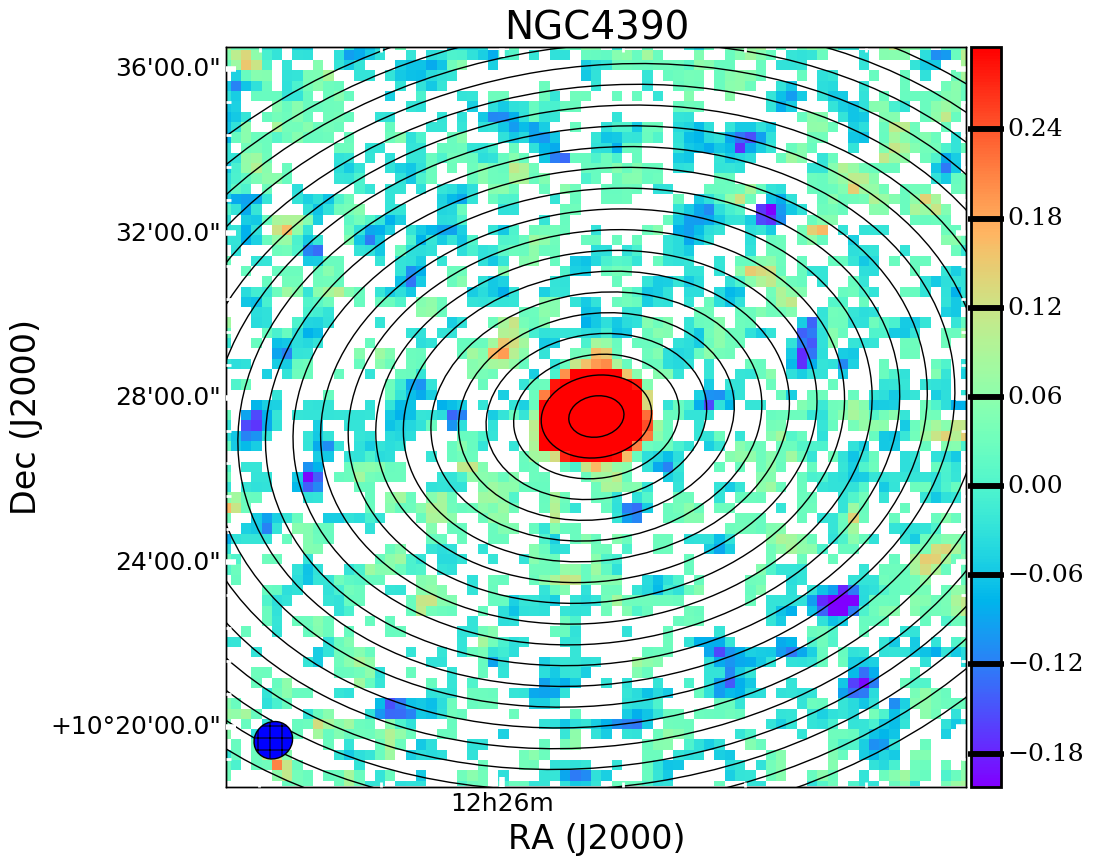}
	\includegraphics[width=5.75cm]{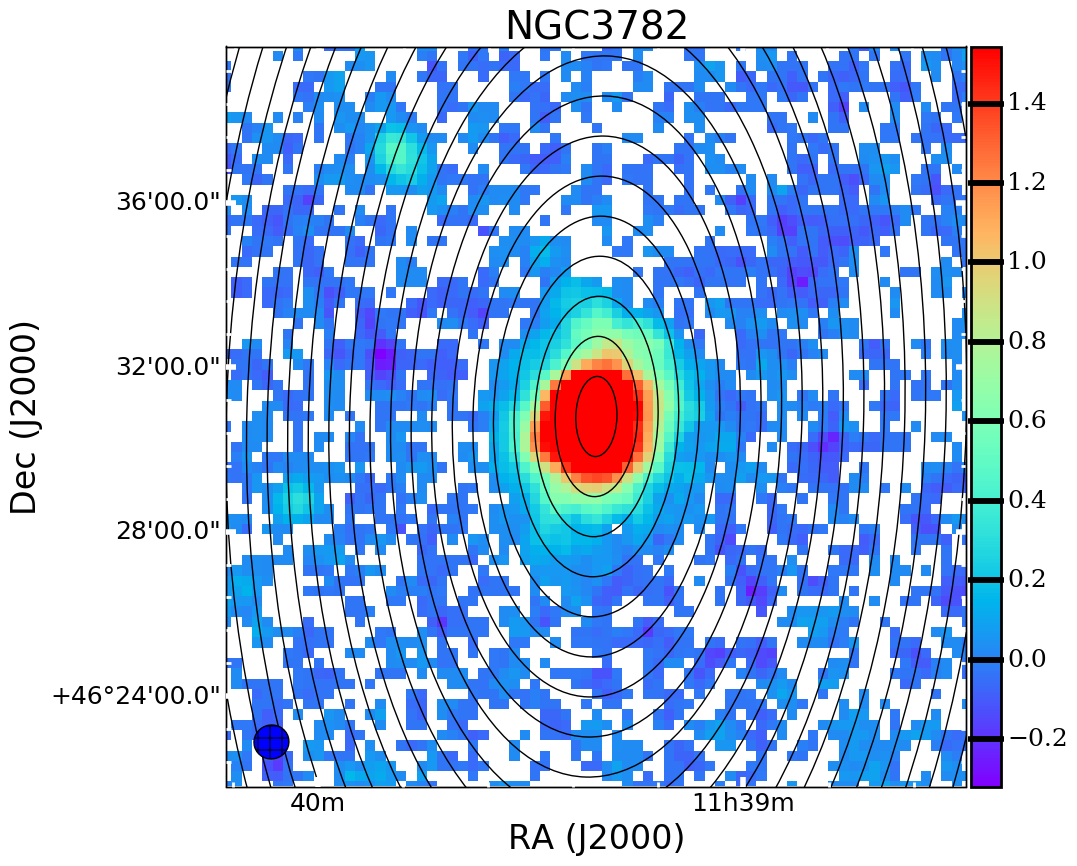}
	\includegraphics[width=5.75cm]{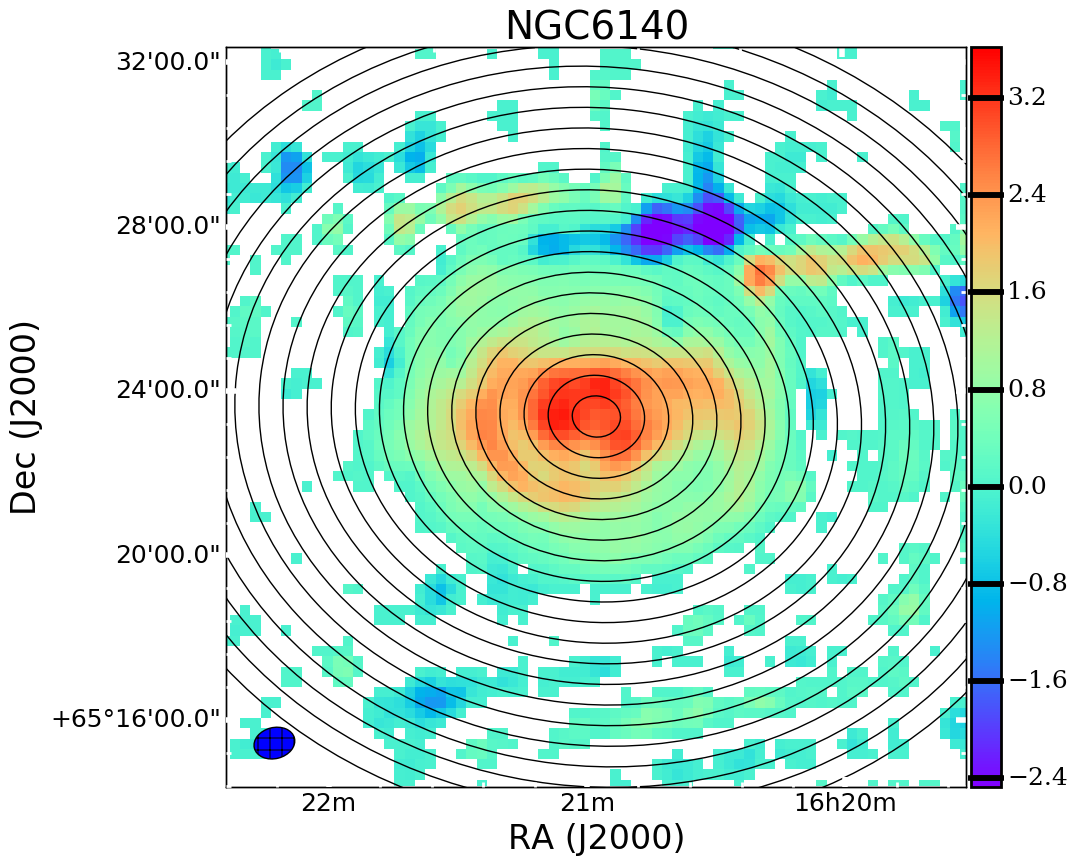}
	\\
	15B-111:\\
	\includegraphics[width=5.75cm]{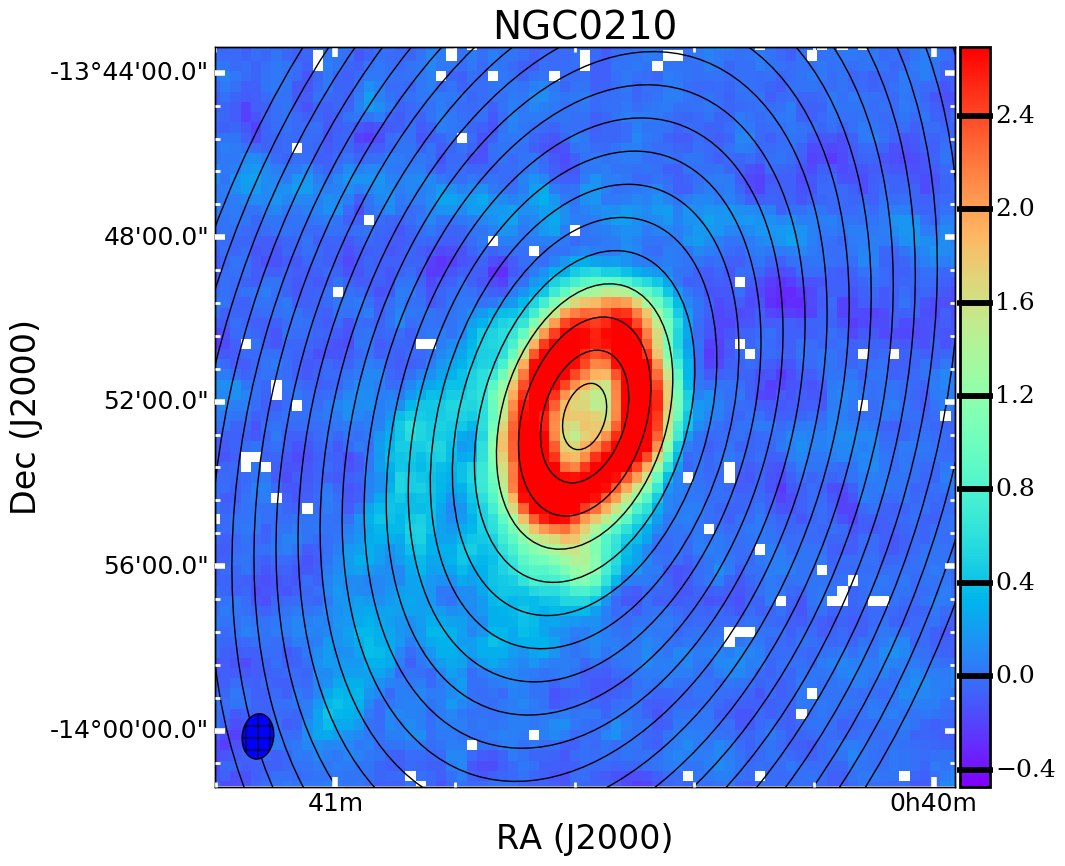}
	\includegraphics[width=5.75cm]{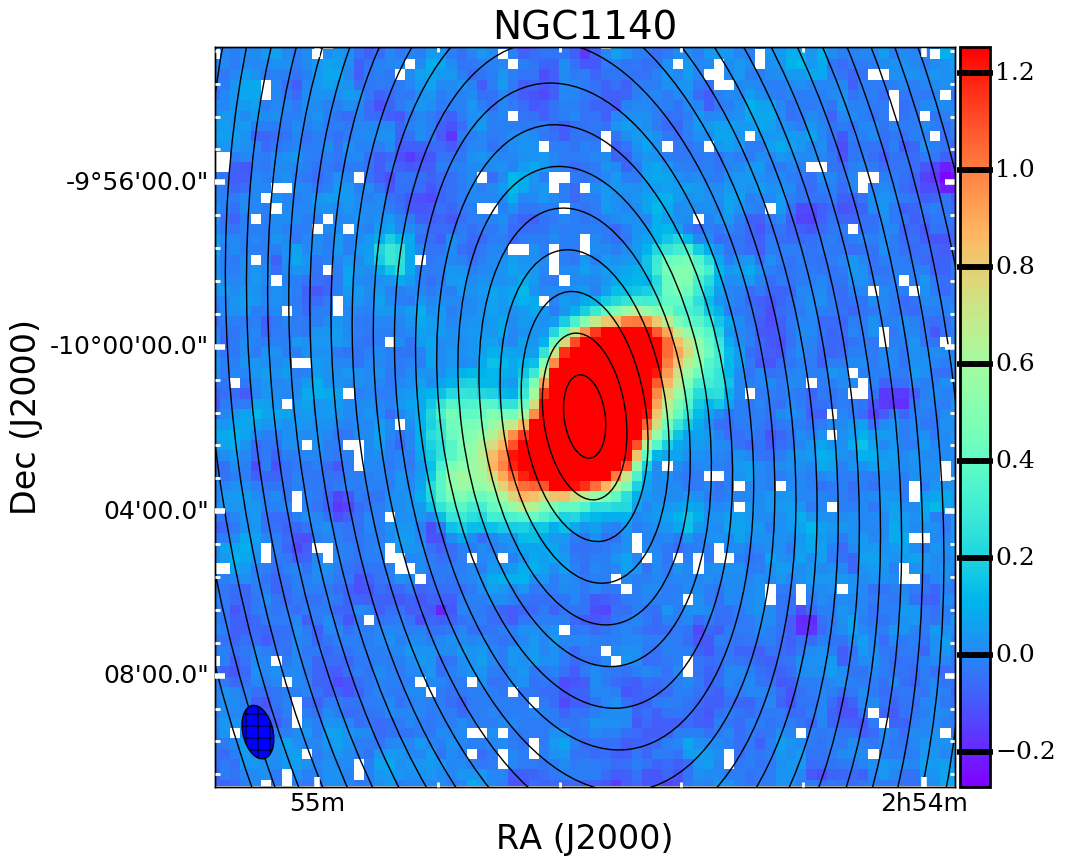}
	\\
	\includegraphics[width=5.75cm]{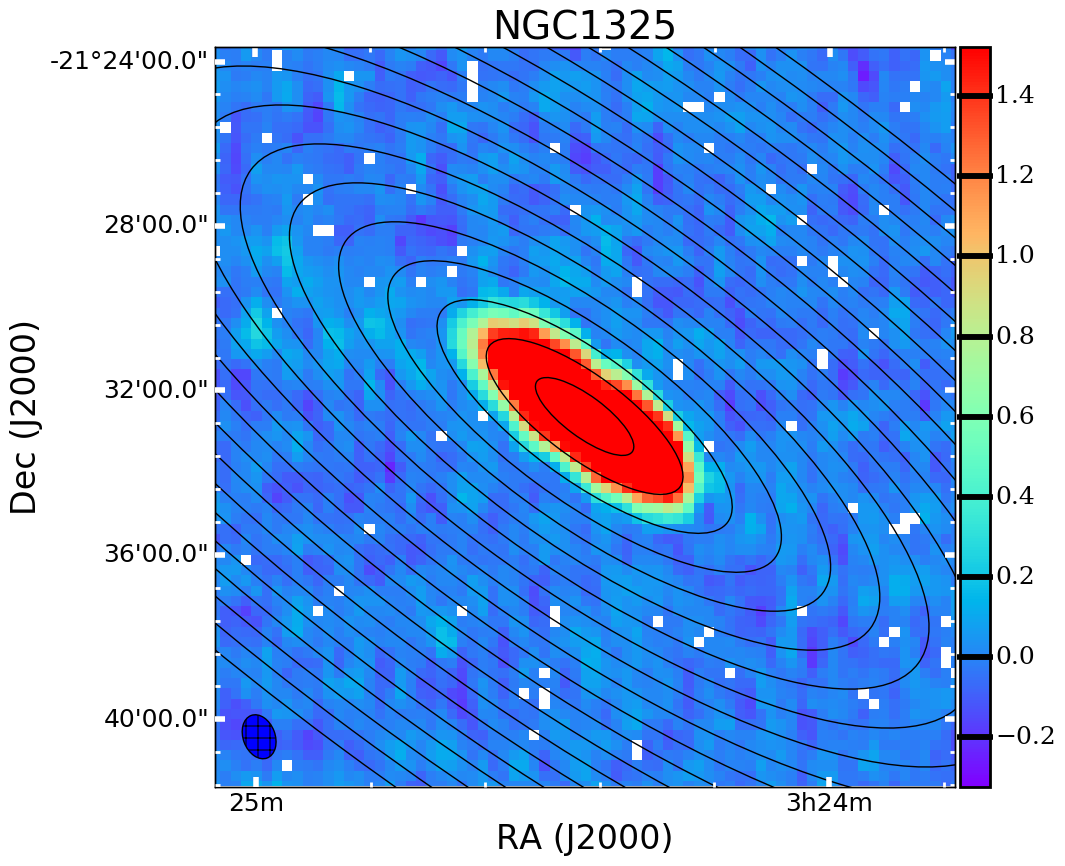}
	\includegraphics[width=5.75cm]{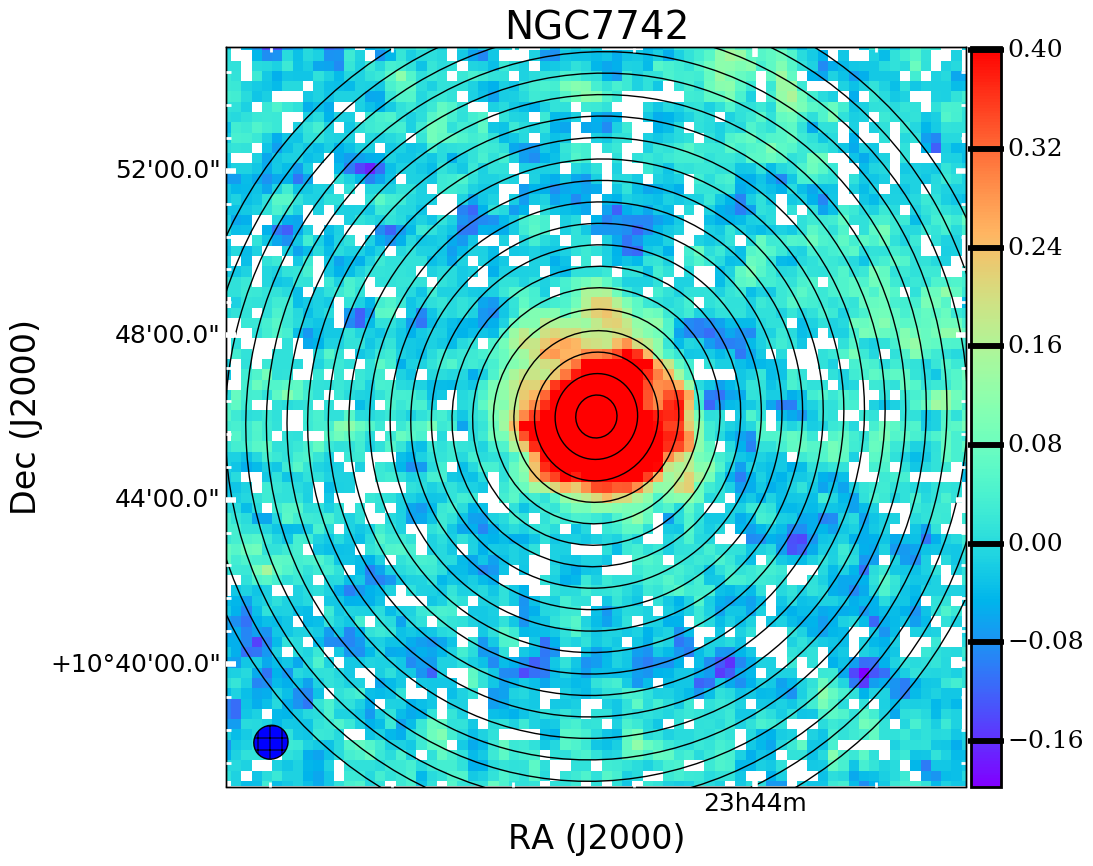}
	\caption{The \hi 21 cm moment zero maps of the galaxies from the AW701 and 15B-111 VLA programs, in units of Jy beam$^{-1}$ km s$^{-1}$. The annuli used for the creation of the radial profiles are overlaid in black and the beam is shown in the lower left corner of each image. The 6 AW701 galaxies are presented first, followed by the 15B-111 galaxies.}\label{fig-himaps}
\end{figure*}

For the \hi\ data, we have collected maps from the VLA for 25 galaxies. First, we use data from the VLA Imaging of Virgo in Atomic Gas (VIVA) survey \citep{Chung09}, where we have downloaded the moment zero maps from their website\footnote{http://www.astro.yale.edu/viva/} for the 14 galaxies that overlap with the NGLS sample. Most of these observations were taken with the VLA in the B- or C-array configuration, with some beam sizes approaching the 15'' resolution of the JCMT. There is also one galaxy, NGC3077, with \hi\ data from the THINGS survey\footnote{http://www.mpia.de/THINGS/Overview.html} \citep{Walter08}. We observed an additional 10 NGLS galaxies with the VLA in the D-array configuration (VLA Project Identifier: AW701, 15B-111). The 6 AW701 galaxies were reduced manually using {\sc CASA} while the newer 4 15B-111 galaxies were reduced using the automated VLA pipeline. To create the integrated \hi\ intensity maps, we also perform a $-1\sigma$ to $1\sigma$ noise cut on the datacube. We present images of the integrated intensity (moment zero) maps for these 10 galaxies in Figure~\ref{fig-himaps}.
\par
For the star formation rate data, we use the \ha\ observations from \citet{SanchezGallego12}, which presented the star formation properties of the galaxies in the NGLS using a combination of archival data and new observations. We use the reduced FITS files, corrected using the R-band continuum filter and the provided conversion factor between \ha\ counts to \ha\ flux to convert the original image into units of the star formation rate. The full procedure to correct the \ha\ maps for contamination from the [N II] lines and the process to estimate the A(H$\alpha$) factor, the internal absorption of H$\alpha$, can be found in \citet{SanchezGallego12}.
\par
We supplement this with \ha\ data from the Herschel Reference Survey \citep{Boselli15}, where we download the \ha\ images from their website\footnote{http://hedam.lam.fr/HRS/}. We also include observations for one galaxy (NGC4273) from the \ha3 survey, taken from the GOLDMINE database \citep{Gavazzi12}. For these galaxies, we first measure the \ha\ counts for the galaxies using an aperture created by eye to capture most of the flux from these galaxies. We combine this information with the published \ha\ flux values to determine the appropriate conversion factor. In order to maintain continuity with the previous papers in this series, we follow the procedure outlined in \citet{SanchezGallego12} to correct the \ha\ flux for [N II] contamination and A(H$\alpha$) factor. For many of these galaxies, we also had to manually align the \ha\ maps, as co-ordinate information was not included in the FITS files. For those cases, we compared to Digital Sky Survey images of the galaxies and fit to star positions using at least 5 stars for each image.
\par
In summary, there are 25 galaxies with \hi\ maps, 33 with \h2\ maps, and 43 with \ha\ maps. More information about the sample can be found in Table~\ref{tab-sample}, where the individual galaxies are listed, along with their available datasets, their inclinations and physical distances. Full tables of the key physical properties of the spiral galaxies in our sample can be found in \citet{Mok16}.
\begin{table*}
	\centering
	\caption{Available \hi, \h2, and \ha\ Data for the Galaxies in Our Sample}\label{tab-sample}
	\begin{tabular}{lcccccccc}
		\hline
		Galaxy\footnotemark[1] & Env\footnotemark[2] & \hi\ source & \hi\ beam size\footnotemark[3] & \hi\ rms\footnotemark[3] & \h2\footnotemark[4] & \ha\ & Incl & Dist \\
		- & - & - & - & [mJy/beam] & - & - & [$\deg$] & [Mpc] \\
		\hline
		{\bf NGC0210} & F & 15B-111 & 66"$\times$46" & 1.75 & D & HRS & 55.4 & 22.4 \\
		NGC3437 & F & - & - & - & D & HRS & 72.8 & 20.1 \\
		NGC6140 & F & AW701 & 60"$\times$45" & 2.16 & ND & NGLS & 32.2 & 17.8 \\
		{\bf NGC7742} & F & 15B-111 & 51"$\times$49" & 1.57 & D & HRS & 16.8 & 24.8 \\
		\hline
		{\bf IC0750} & G & AW701 & 51"$\times$49" & 1.30 & D & NGLS & 65.8 & 8.3 \\
		IC3908 & G & - & - & - & D & HRS & 73.2 & 19.0 \\
		NGC0450 & G & - & - & - & D & NGLS & 49.8 & 25.4 \\
		NGC1140 & G & 15B-111 & 79"$\times$44" & 1.51 & ND & NGLS & 73.8 & 20.1 \\	
		NGC1325 & G & 15B-111 & 66"$\times$47" & 1.74 & ND & NGLS & 74.3 & 20.7 \\		
		NGC2146A & G & AW701 & 63"$\times$46" & 1.50 & ND & NGLS & 69 & 25.9 \\
		{\bf NGC2742} & G & AW701 & 62"$\times$45" & 1.24 & D & NGLS & 60.6 & 22.4 \\
		{\bf NGC3077} & G & THINGS & 14"$\times$13" & 0.94 & D & NGLS & 38.1 & 3.9 \\
		NGC3162 & G & - & - & - & D & NGLS & 37.1 & 20.7 \\
		NGC3227 & G & - & - & - & D & NGLS & 68.3 & 18.4 \\
		NGC3346 & G & - & - & - & D & HRS & 34.1 & 19.5 \\
		NGC3507 & G & - & - & - & D & NGLS & 31.9 & 15.5 \\
		NGC3684 & G & - & - & - & D & HRS & 50.8 & 18.4 \\
		NGC3782 & G & AW701 & 51"$\times$49" & 1.26 & ND & NGLS & 60.3 & 14.3 \\
		NGC3982 & G & - & - & - & D & HRS & 29.9 & 20.1 \\
		NGC4041 & G & - & - & - & D & NGLS & 22 & 21.9 \\
		NGC4123 & G & - & - & - & D & HRS & 44.3 & 20.1 \\
		{\bf NGC4713} & G & VIVA & 26"$\times$22" & 1.96 & D & HRS & 23.8 & 10.9 \\
		NGC4771 & G & - & - & - & D & HRS & 74.4 & 17.2 \\
		NGC4772 & G & VIVA & 18"$\times$15" & 0.36 & ND & NGLS & 67.3 & 16.1 \\
		NGC4775 & G & - & - & - & D & HRS & 28.4 & 23.0 \\
		{\bf NGC4808} & G & VIVA & 40"$\times$36" & 0.59 & D & HRS & 69.2 & 12.0 \\
		\hline		
		{\bf NGC4254} & V & VIVA & 38"$\times$33" & 0.41 & D & NGLS & 20.1 & 16.7 \\
		NGC4273 & V & - & - & - & D & GOLDMINE & 26.9 & 16.7 \\
		NGC4294 & V & VIVA  & 29"$\times$27" & 0.29 & ND & HRS & 70.2 & 16.7 \\
		{\bf NGC4298} & V & VIVA & 17"$\times$16" & 0.35 & D & NGLS & 58.4 & 16.7 \\
		NGC4303 & V & - & - & - & D & HRS & 18.1 & 16.7 \\
		{\bf NGC4383} & V & VIVA & 45"$\times$38" & 0.26 & D & NGLS & 63.7 & 16.7 \\
		NGC4390 & V & AW701 & 58"$\times$53" & 1.16 & ND & NGLS & 43.3 & 16.7 \\
		NGC4396 & V & VIVA & 27"$\times$27" & 0.28 & ND & HRS & 71.6 & 16.7 \\
		NGC4430 & V & - & - & - & D & NGLS & 43.8 & 16.7 \\
		NGC4480 & V & - & - & - & D & HRS & 61.2 & 16.7 \\
		NGC4548 & V & VIVA & 17"$\times$16" & 0.30 & ND & HRS & 36.9 & 16.7 \\
		{\bf NGC4567} & V & VIVA & 17"$\times$16" & 0.36 & D & NGLS & 39.4 & 16.7 \\
		{\bf NGC4568} & V & VIVA & 17"$\times$16" & 0.36 & D & NGLS & 67.5 & 16.7 \\
		{\bf NGC4579} & V & VIVA & 42"$\times$35" & 0.45 & D & NGLS & 41.9 & 16.7 \\
		NGC4647 & V & - & - & - & D & NGLS & 31.6 & 16.7 \\
		{\bf NGC4651} & V & VIVA & 17"$\times$16" & 0.40 & D & NGLS & 49.5 & 16.7 \\
		{\bf NGC4654} & V & VIVA & 16"$\times$16" & 0.45 & D & HRS & 59.8 & 16.7 \\
	\hline
	\end{tabular}
	\begin{tabular}{l}
		\footnotemark[1] Galaxies with both \hi\ and \h2\ data are in bold face.\\
		\footnotemark[2] For the environment, F indicates field galaxies, G indicates group galaxies, and V indicates Virgo galaxies. \\ A full explanation
		of our categories can be found in \cite{Mok16} \\
		\footnotemark[3] Beam sizes and rms from the \hi\ maps in the VIVA survey are from Table 2 of \citet{Chung09}\\
		\footnotemark[4] For \h2\ data, D indicates detected galaxies, ND indicates non-detected galaxies \\
	\end{tabular}
\end{table*}
\subsection{Integrated Properties of the Resolved Sample}
\subsubsection{Comparison with the Remaining NGLS Spiral Galaxies}
\par
First, we compare our resolved subset of 43 galaxies to the remaining sample of 55 NGLS spiral galaxies from \citet{Mok16} that were not included in this paper, as they did not possess detected \co\ emission or VLA \hi\ observations or they have high inclinations. We find that the two samples have similar \hi\ gas masses (log M$_{\rm H{_I}} = 9.16\pm0.05$ for the resolved sample vs. log M$_{\rm H{_I}} =  9.05\pm0.04$ for the non-included sample). The similar total \hi\ gas mass is likely due to the \hi\ flux selection imposed by the NGLS. We also find no significant difference between the integrated \hi\ properties of the Virgo galaxies in the resolved sample, compared to the Virgo galaxies that were not included in this paper. These results indicate that we are not selecting a potentially biased \hi\ sample out of the original \hi-flux selected sample.
\par
On the other hand, the stellar masses for the resolved sample (log M$_{*} = 9.94\pm0.06$) are significantly higher than the non-included sample (log M$_{*} = 9.37\pm0.07$). In \citet{Mok16}, we found that the stellar mass for the \co\ detected galaxies is significantly higher than the \co\ non-detected galaxies. Thus, it is not surprising to see a similar difference in stellar mass, since this resolved sample would comprise most of the \co\ detected galaxies, except for a few heavily inclined galaxies. In addition, the star formation rates are higher in the resolved sample. This is likely due to the strong link between \h2\ gas and star formation, as well as between stellar mass and star formation. These results suggest that the resolved sample would not contain many of the gas-rich low surface brightness galaxies discussed in \citet{Mok16} and in other surveys of these \hi-rich objects, such as HIghMass \citep{Huang14}.
\subsubsection{Comparison of the Virgo and non-Virgo Galaxies in the Resolved Sample}
\par
For our resolved sample, we present a summary of the global properties of the non-Virgo and Virgo galaxies in Table~\ref{tab-ressample2}. In this analysis, we have combined the field and group galaxy samples to increase the number of galaxies in the non-Virgo sample.
\par
Although the non-Virgo galaxies have higher \hi\ masses, the difference in the stellar mass normalized \hi\ masses is less significant, as shown by performing the \ks test on the two distributions. While Virgo galaxies have a higher average stellar mass, it is less than a 1$\sigma$ difference and the \ks test shows that the two distributions cannot be distinguished. The other properties, such as star formation rate and specific star formation rate (sSFR), are also comparable between the two samples. These results give us more confidence that we are comparing between similar galaxies and probing the more subtle effects of environment.
\par
The main difference between the Virgo and non-Virgo resolved samples is in the molecular gas properties. Since not all of these galaxies are detected in the \co\ maps, as shown in Table~\ref{tab-sample}, we have used the statistical technique of survival analysis on these two samples, as discussed in \citet{Mok16}. The resolved sample shows the same general trends as analysis from the full NGLS sample, with the Virgo galaxies possessing higher mean molecular gas masses, higher \h2-to-\hi\ ratios, and lower star formation efficiencies (or longer molecular gas depletion times). For the total gas mass, the two samples are quite similar, suggesting that the cluster environment may be aiding in the conversion process between atomic and molecular gas. We will further explore these global results by measuring the radial profiles of these galaxies.
\begin{table}
	\begin{minipage}{120mm}
	\caption{Global Properties of the non-Virgo and Virgo Resolved Samples}\label{tab-ressample2}
	\begin{tabular}{lcccc}
	\hline
	Mean Quantity & Non-Virgo & Virgo & KS - Test \\
	 & ($26$) & ($17$) &\\
	\hline
	log M$_{\rm H_{I}}$ [M$_\odot$] & $9.24\pm0.05$ & $9.04\pm0.08$& $\underline{0.009}$\\
	log M$_{*}$ [M$_\odot$] & $9.87\pm0.07$ & $10.05\pm0.12$ & $0.154$ \\
	M$_{\rm H_{I}}$/M$_{*}$ & $0.35\pm0.06$ & $0.15\pm0.03$ & $0.108$\\
	Distance [kpc] & $18.6\pm0.94$ & $16.0$ & - \\
	\hline
	log SFR [M$_\odot$ yr$^{-1}$] & $-0.29\pm0.09$ & $-0.25\pm0.14$ & $0.774$\\
	log sSFR [yr$^{-1}$] & $-10.17\pm0.09$ & $-10.30\pm0.10$ & $0.610$\\
	log M$_{\rm H_{I}}$/SFR [yr] & $9.54\pm0.07$ & $9.29\pm0.11$ & $0.486$\\
	\hline
	log M$_{\rm H_2}$\footnotemark[1] [M$_\odot$] & $8.35\pm0.11$ & $8.83\pm0.13$ & $\underline{0.028}$\footnotemark[2] \\
	log SFR/M$_{\rm H_2}$\footnotemark[1] [yr] & $-8.62\pm0.11$ & $-9.06\pm0.10$ & $\underline{0.049}$\footnotemark[2] \\
	M$_{\rm H_{2}}$/M$_{*}$\footnotemark[1] & $0.047\pm0.012$ & $0.095\pm0.016$ & $0.051$\footnotemark[2] \\
	M$_{\rm H_{2}}$/M$_{\rm H_{I}}$\footnotemark[1] & $0.38\pm0.14$ & $1.43\pm0.37$ & $\underline{0.007}$\footnotemark[2] \\
	M$_{\rm H_{2}}$ + M$_{\rm H_{I}}$ [M$_\odot$]\footnotemark[1] & $9.35\pm0.05$ & $9.35\pm0.10$ & $0.656$\footnotemark[2] \\
	\hline
	\end{tabular}
	\begin{tabular}{l}
	Note: Underline indicates $p<0.05$. \\
	\footnotemark[1] CO non-detections taken into account using survival analysis; \\
	please refer to \citet{Mok16} for more information. \\
	\footnotemark[2] Log-rank test used, which takes into account censored data. \\
	\end{tabular}
	\end{minipage}
\end{table}
\subsection{Creating Surface Density Maps}
\par
We convert the HARP \co\ moment maps to a \h2\ surface density map by assuming a CO-to-\h2\ conversion factor of $X_{\rm CO} = 2\times10^{20} {\rm cm}^{-2} ({\rm K\ km\ s}^{-1})^{-1}$ \citep{Strong88} or $\alpha_{\rm CO} = 3.2\ {\rm M}_\odot\ {\rm pc}^{-2}({\rm K\ km\ s}^{-1})^{-1}$. We use a constant line ratio (R$_{31}$) between \co\ and CO $J=1-0$ of 0.18, the average value found for galaxies in the NGLS \citep{Wilson12}. This leads to the following relation between $\Sigma_{\rm H_2}$, the surface density of molecular hydrogen, and $I_{\rm CO(3-2)}$, the integrated \co\ intensity:
	\begin{equation}
	\Sigma_{\rm H_2} [{\rm M}_\odot\ {\rm pc}^{-2}] = 17.8\times({\rm R}_{31}/0.18)^{-1}I_{\rm CO(3-2)} [{\rm K\ km\ s} ^{-1}]
	\end{equation}
\par
To convert the VLA 21 cm moment zero maps into physical units, we combine equations 1 and 5 from \citet{Walter08} and then convert to units of M$_\odot$ pc$^{-2}$. This leads to the following relationship, where $\Sigma_{\rm HI}$ is the surface density of \hi, $\sum S \Delta \nu$ is the integrated flux density from the moment zero maps, and FWHM$_{\rm maj}\times$ FWHM$_{\rm min}$ is a measure of the beam area:
	\begin{equation}
	\Sigma_{\rm HI} [{\rm M}_\odot\ {\rm pc}^{-2}] = 8765.27\times\frac{\sum S \Delta \nu [{\rm Jy\ beam ^{-1} km\ s}^{-1}]}{{\rm FWHM_{\rm maj}}\times{\rm FWHM_{\rm min}}}\
	\end{equation}
\par
For the \ha\ maps, we convert the \ha\ counts into star formation rates using the procedure outlined in \S~\ref{sec-dataselection}. This process uses the conversion factor from \citet{Kennicutt09}, which assumes a Kroupa IMF \citep{Kroupa03}:
	\begin{equation}
	{\rm SFR} [{\rm M}_\odot\ {\rm yr}^{-1}] = 5.5 \times 10^{-42} \, {\rm L(H\alpha) [erg\ s^{-1}]}
	\end{equation}
where SFR is the star formation rate and L(H$\alpha$) is the H$\alpha$ luminosity. When we generate radial profiles, we divide the star formation rates in each measured annulus by its physical size, as determined by the distance to each galaxy listed in Table~\ref{tab-sample}. This results in a final measurement of the surface density of star formation ($\Sigma_{\rm SFR}$) in units of M$_\odot$ yr$^{-1}$ pc$^{-2}$.
\par
We perform an additional data reduction step on the \co\ and \ha\ data. Our VLA dataset has a wide variety of angular resolutions while our \co\ dataset has a constant angular resolution of 15''. To allow for analysis using a combination of two values (such as the \h2/\hi\ ratio) and to better compare between the \h2\ and \hi\ datasets, we adopted a standard 60'' resolution. For the galaxies observed using the D-array configuration, we convolved the \co\ and \ha\ maps to the same resolution as the \hi\ maps using the software package {\sc Starlink}. For the VIVA dataset, observed using the VLA in the B/C configurations, we convolved all three maps (\h2, \hi, and \ha) to our standard 60'' resolution.
\par
Finally, we correct for any inclination effects by multiplying our resulting maps with a $\cos i$ factor, with the inclinations provided from the HyperLeda database \citep{Paturel03}.
\subsection{Generating Radial Profiles}
\par
To generate the radial profiles for our galaxies, we use the co-ordinates, position angle, and inclination data from the HyperLeda database to define 20 concentric annuli. The widths of each ring along the minor axis is set to 30'', which is half of the standard angular resolution for our dataset.  We then measured the mean values within each aperture using the {\sc apperadd} command. For galaxy pairs that were collected in a single map, such as NGC4567 and NGC4568, the other galaxy was masked out manually from the individual maps. We note that the radial profiles will only extend as far out as the observed area in each map and may not include all 20 annuli, especially for the \h2 data.
\par
Furthermore, an annulus is only included if it contains more than 5$\%$ valid (non-blank) pixels. We have tested more conservative thresholds, such as 25$\%$, and it did not result in any significant difference in our profiles and results. However, it did reduce the amount of galaxies in our \h2\ sample due to the localized nature of some of our \co\ detections. To maximize the number of galaxies in our sample, we have maintained this low threshold.
\par
To determine any systematic difference between the galaxies in the sample, we divide the physical radii by each galaxy's R$_{25}$, which is half of their D$_{25}$ value from the HyperLeda database. This normalization is important because the Virgo galaxies in our sample are on average slightly closer and physically larger than the non-Virgo sample \citep{Mok16}. A comparison using physical radii (kpc) may be affected by the differences in the size of the galaxies in the Virgo and non-Virgo samples and their distances.
\par
Next, we tested three ways of combining the profiles to create the `average' radial profile for our subsamples: the median, the mean, and the geometric mean. The geometric mean is the square root of the product of the input values. This is also equivalent to the log-average, where we perform the arithmetic mean on the logarthmic values.
\par
\begin{figure*}
	\centering
	\includegraphics[width=7.5cm]{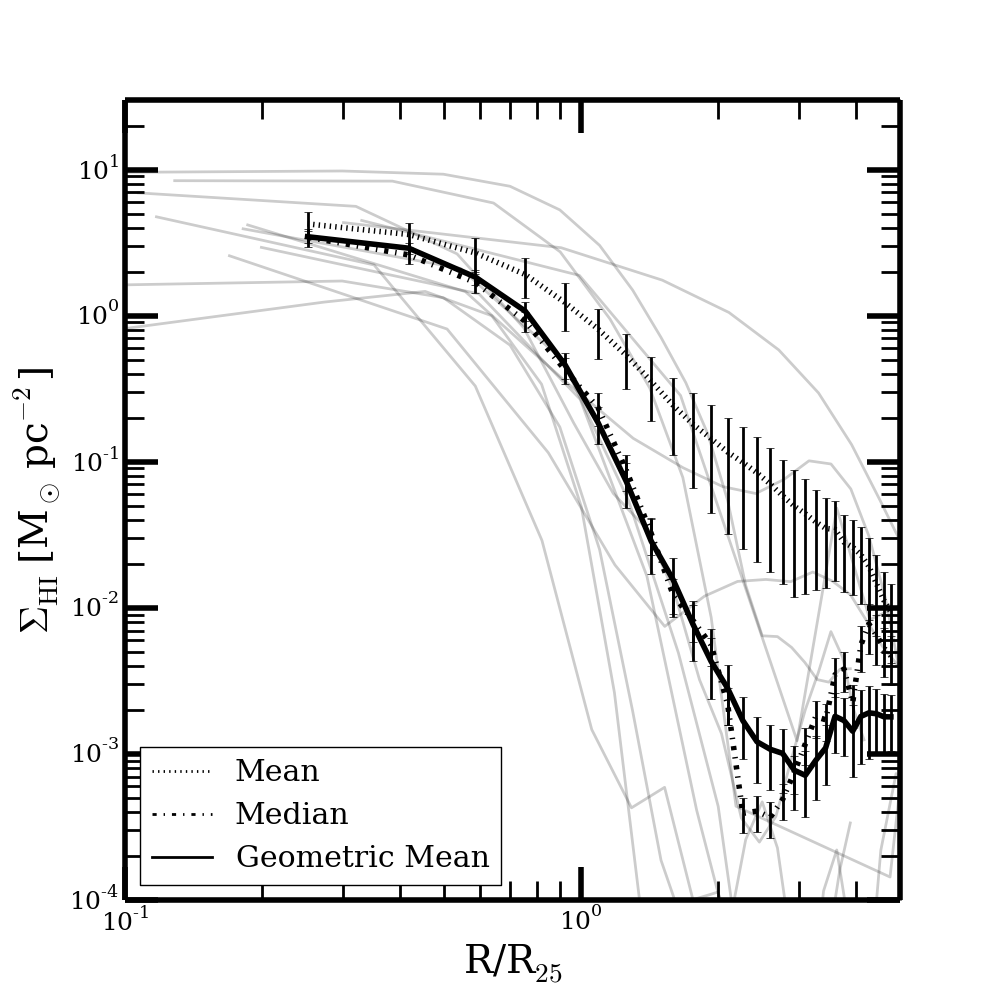}
	\includegraphics[width=7.5cm]{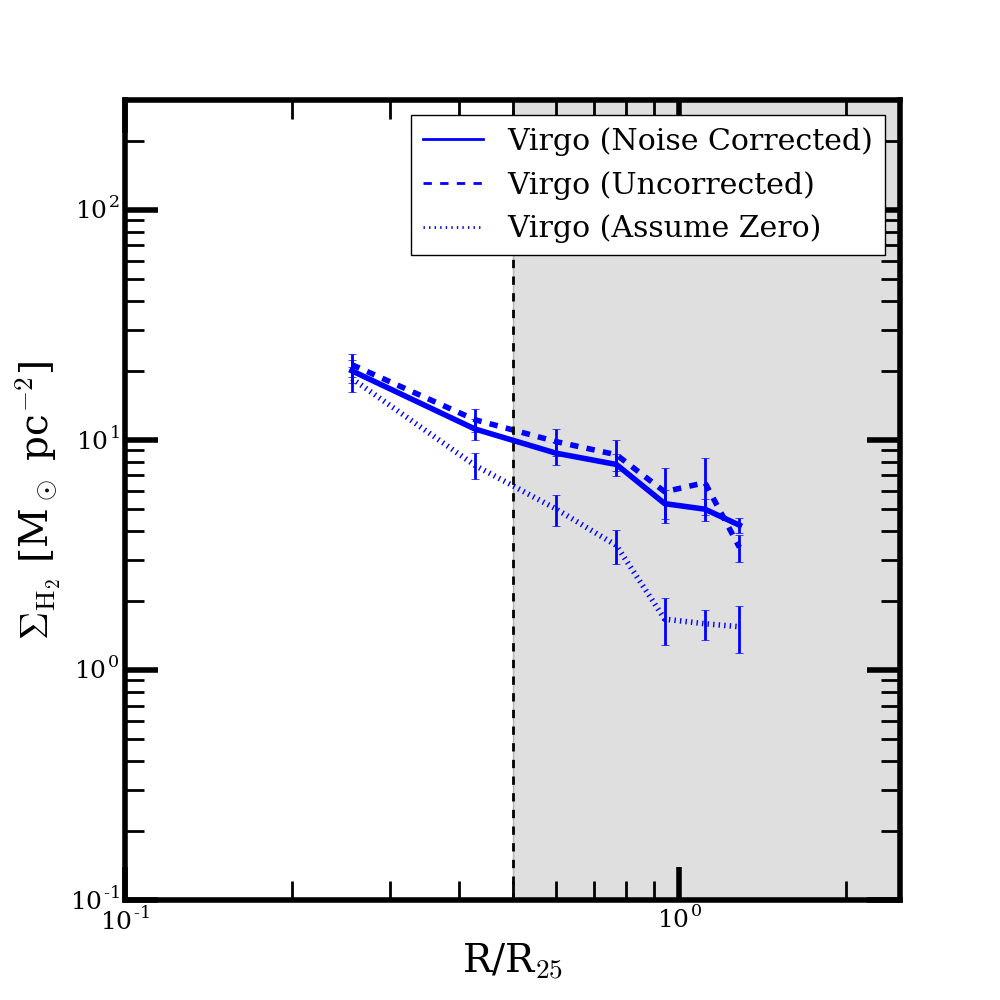}
	\caption{{\bf Left:} A comparison of the results from using different methods for defining the radial profile for the population of Virgo galaxies, including mean (dotted), geometric mean (solid), and median (dot-dashed). Error bars for the mean are statistical. Error bars for the median method are based on median absolute deviation. Error bars for the geometric mean method are based on the formula from \citet{Norris40}. We see that the mean profile is biased by high outliers, while the median method produces jagged profiles in regions due to a small number of galaxies. {\bf Right:} The average \h2\ surface density radial profiles, measured using the geometric mean method, for the Virgo galaxies in our sample. The radii are normalized by R$_{25}$. The dashed vertical line is at R$_{25}$ = 0.5, the main target area of the NGLS survey, and results from the \co\ maps are less reliable in the shaded region. The original, uncorrected profile (dashed line), noise corrected profile (solid line) and the profile assuming blank pixels have no gas (dotted line) is plotted. The three methods produce similar results in the reliable region near the centre, where most of the \co\ pixels are detected.}\label{fig-methodtest}
\end{figure*}

We calculate the average profile at equally spaced points in the R$_{25}$-normalized radius, interpolating between the physical annuli points where necessary. The results of these three methods for the \hi\ data are presented in the left plot of Figure~\ref{fig-methodtest} for the Virgo sample. These averages are only calculated in regions where there are more than three galaxies. The differences in the horizontal range that each radial profile spans depends on the number of galaxies in each sample and the interaction between the fixed size annuli used for the analysis and the distance to each individual galaxy.
\par
The error bars on the mean are the statistical errors and the error bars on the median are based on the median absolute deviation. The error on the geometric mean uses the formula from \citet{Norris40}:
	\begin{equation}
S_G = G \frac{s_{\log x_i}}{\sqrt{n-1}}
	\end{equation}
where $G$ is the geometric mean, $s_{\log x_i}$ is the standard deviation of the log values, and n is the number of galaxies. Since the stellar masses of the Virgo and non-Virgo samples are not significantly different and we are normalizing their radii using R$_{25}$, we can compare the standard error on the mean values of these similar objects. Note that the datapoints where we calculate the average profile will correspond to different points at the R/R$_{25}$ for each individual galaxy, given their varying distances and sizes. Thus, we perform interpolation between the measured annuli points for the individual galaxy profiles when necessary. It is also possible for the error bars to be underestimated, because each data point in average profile is not fully independent of the others. On the other hand, this is compensated by the large number of individual galaxy profiles which goes into the averaging process.
\par
We chose the geometric mean method to generate the average radial profile for our analysis. Compared to the median method, it provides a more stable profile. The median can produce ragged profiles, particularly at radii with few individual measurements. Compared to the mean method, the geometric mean is more representative. The simple mean can be biased by high outliers, especially when presented in a log-log plot. The choice of the particular averaging method does not change the overall results presented in this paper, especially between the median and the geometric mean methods.
\par
Finally, we calculate the average R/R$_{25}$ value for the galaxies in each average radial profile. To create a balance between the increased dataset available from combining our individual profiles and the angular resolution of our data, we decided to generate average radial profiles points at intervals which are the equivalent of 15'' at the average R/R$_{25}$ value for the galaxies making up the average profile.
\subsection{Correcting Blank Pixels in \h2\ Maps}
\par
The \co\ maps for many of our galaxies contain large areas of blank pixels. This is due to the sensitivity limits of the observations and the data reduction procedure from \citet{Wilson12}, which removes regions of the map below a certain signal to noise limit. One approach is to assume that the non-detected pixels have no molecular gas present. This correction is performed by multiplying the raw value by the ratio of valid pixels to the total number pixels in the aperture. This can be considered a lower limit to the \h2\ profile. Another approach is to add in the average 1$\sigma$ noise of the map for those missing pixels. This second correction method assumes that the undetected pixels would take on the mean value from the noise map generated in the data reduction process, where a line width of 100 km/s is assumed. The final approach is to make no corrections to the average value of the detected pixels measured in the aperture, which we will call the uncorrected profile.
\par
We compare these three methods to correct for these blank pixels in the right plot of Figure~\ref{fig-methodtest}, looking at the \h2\ radial profiles for our Virgo sample. These corrections are not very important within R/R$_{25} = 0.5$, the main target area of the NGLS survey. Inside this region, most of the pixels are detected and the three curves are similar to each other. We decided to apply the noise correction method to all of the subsequent \h2\ maps in our sample, as this is the most physical scenario. In general, the noise corrected profile provides similar results to the uncorrected case. The radial profile for the case where we assume no molecular gas is present in the blank pixels naturally produces steeper profiles, as it reduces the surface density in the outskirts of these galaxies.
%
%
%
\section{Results and Discussion}
\subsection{\hi\ Radial Profiles - Stripping in Group and Virgo Galaxies}\label{sec-res_hiprofiles}
\par
\begin{figure*}
	\centering
	\includegraphics[width=7.5cm]{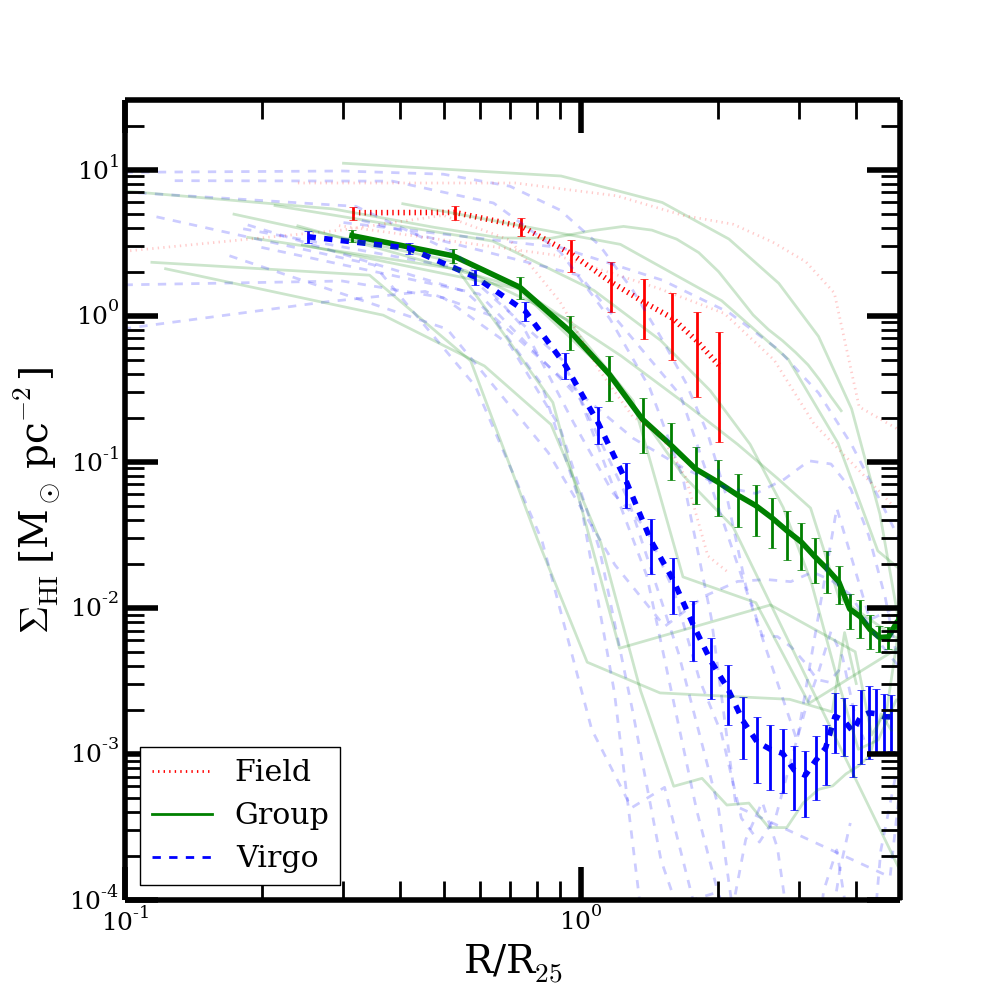}
	\includegraphics[width=7.5cm]{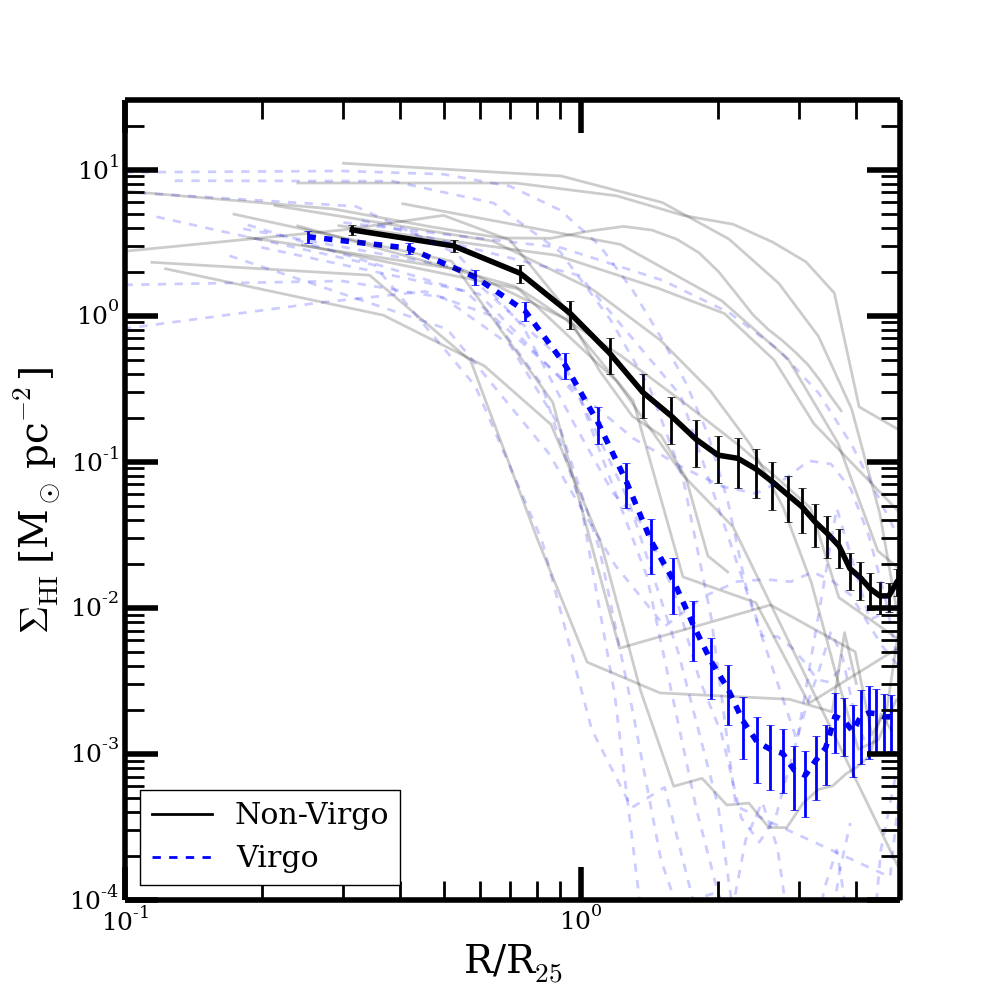}
	\caption{The radial profiles of \hi surface density for the galaxies in our sample, calculated using the geometric mean method and normalized by R$_{25}$. The statistical error bars for each profile are plotted. The individual galaxy profiles are in the background. {\bf Left:} The sample is separated into field (red-dotted), group (green), and Virgo (blue-dashed) populations. We see a reduction in the sizes of the \hi\ disk, from field to group to the Virgo Cluster, suggesting \hi\ properties are affected even in moderate density environments. {\bf Right:} The sample is separated into the non-Virgo (black) and Virgo (blue-dashed) populations. Even for this \hi-flux selected sample, the Virgo Cluster galaxies have truncated \hi\ distributions in the outskirts compared to non-Virgo galaxies.}\label{fig-hi}
\end{figure*}

We present the radial profiles for our sample of galaxies with available \hi\ data in Figure~\ref{fig-hi}. We separate the sample of galaxies by field, group, and Virgo samples and also between the Virgo and non-Virgo samples. All of the radial profiles have a relatively flat inner portion, which transitions into a steeper profile in the outskirts. We note that the galaxies in the Virgo cluster have truncated \hi\ disks compared to non-Virgo galaxies, which was seen in previous studies of Virgo Cluster spirals \citep{Cayatte94}. The new result here is that this truncation is seen even in this sample of \hi-flux selected galaxies from the NGLS.
\par
Comparing the smaller field and group samples, the field galaxies have a shallower decrease in the outskirts compared to the group galaxies. This would suggest a natural trend in \hi\ disk sizes from the field to the group and then to the cluster environment. Given we only have 3 field galaxies with \hi\ maps in our sample, we may need more observations to confirm this difference with the group sample. On the other hand, this scenario fits well with the picture of ram-pressure stripping, where higher density environments can strip away the ISM of spiral galaxies. Our result provides some more evidence for \hi\ stripping in environments of relatively modest density. Observationally, \hi-deficient galaxies have been found in the group environment, such as in Hickson Compact Groups \citep{Martinez-Badenes12}. The Blind Ultra Deep \hi\ Environmental Survey (BUDHIES) have also found a correlation between the fraction of \hi\ detected galaxies and the mass of their host groups \citep{Jaffe16}.
\par
In addition, since this is a \hi-flux selected sample from the NGLS, these galaxies still possess a relatively large amount of available cold gas. It is likely that the NGLS is preferentially selecting galaxies that are recently infalling into the Virgo Cluster or have not been strongly affected by the cluster environment. In addition, due to the \hi-flux criteria and with most of the galaxies in the resolved sample detected in the \co\ maps, we are selecting relatively high stellar mass galaxies that are less likely to be fully stripped according to semi-analytical models \citep{Luo16}. Despite these caveats, the trends observed in the \hi\ radial profile suggest that the environment can still play a big factor in changing the spatial distribution of the ISM inside these galaxies.
\begin{figure*}
	\centering
	\includegraphics[width=7.5cm]{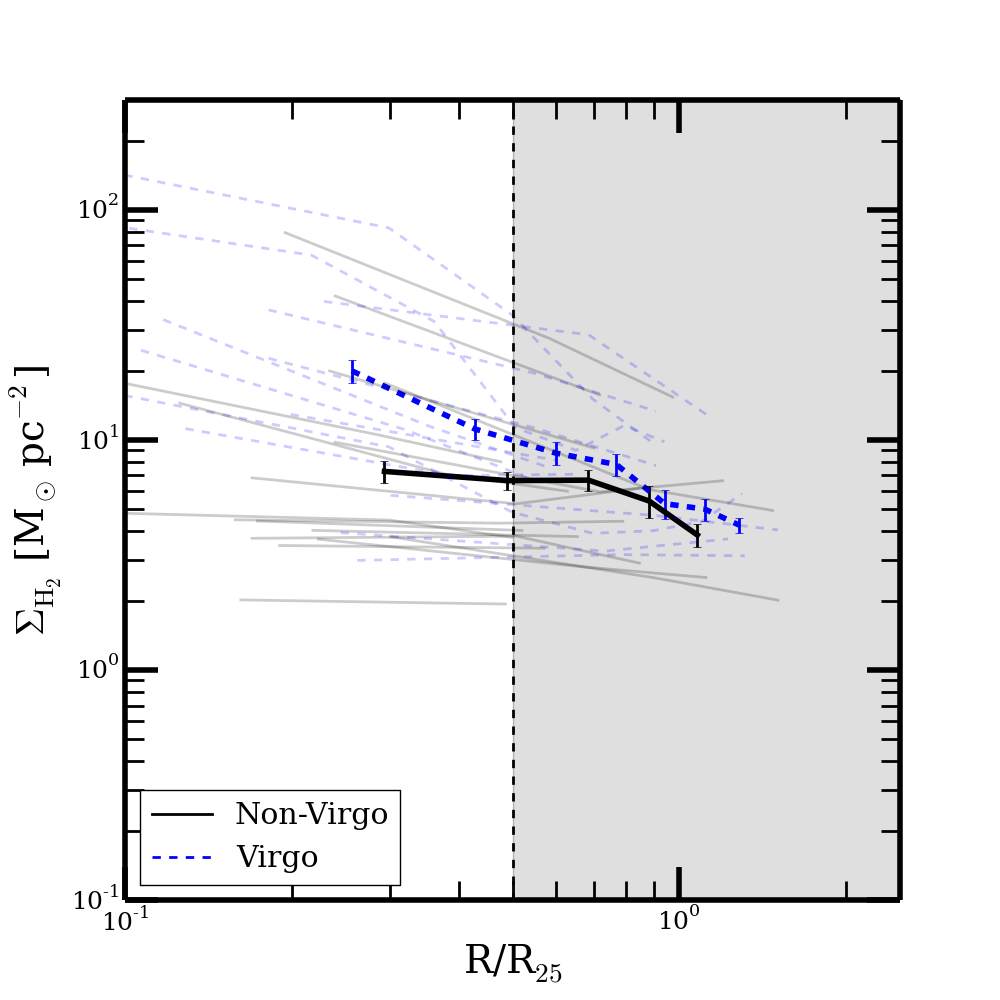}
	\includegraphics[width=7.5cm]{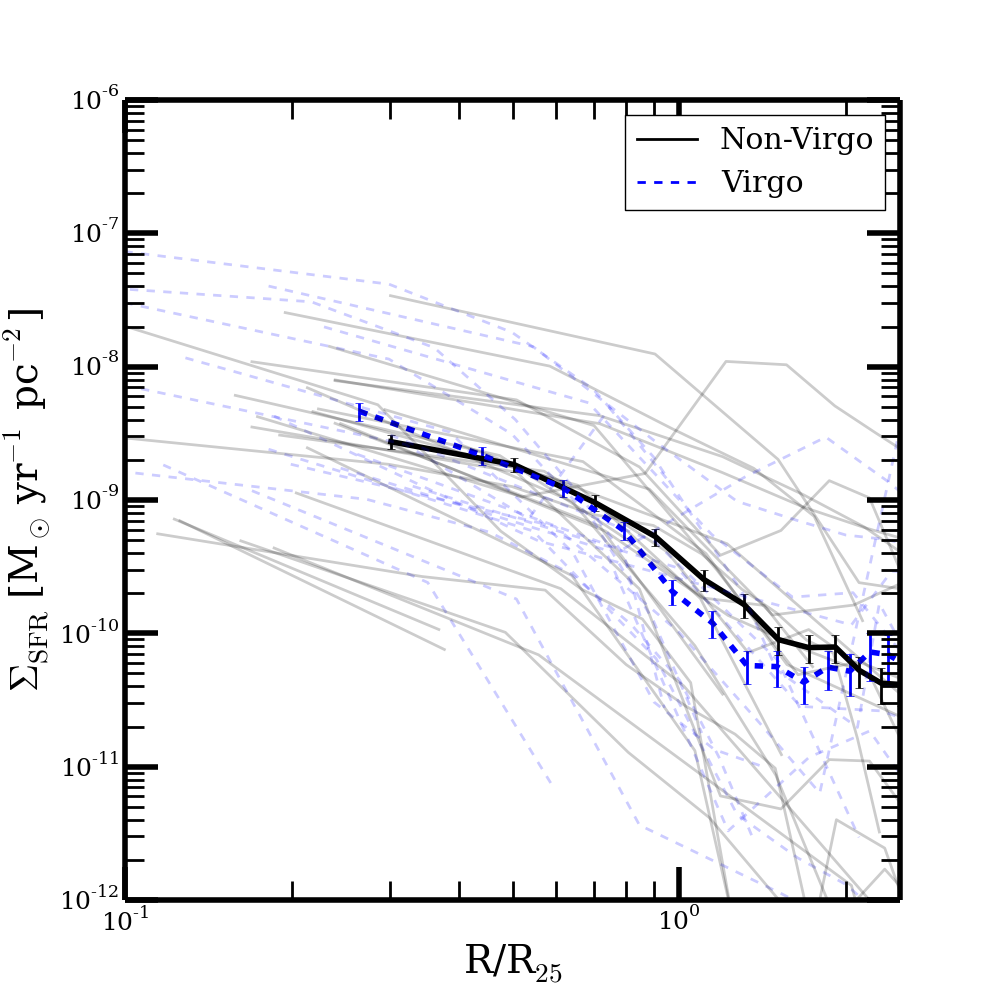}
	\caption{{\bf Left:} The radial profiles of \h2 surface density for the galaxies in our sample, calculated using the geometric mean method and normalized by R$_{25}$. The statistical error bars for each profile are plotted. The individual galaxy profiles are in the background. The sample is separated into the non-Virgo (black) and Virgo (blue-dashed) populations. The dashed vertical line is at R$_{25}$ = 0.5, the main target area of the NGLS survey, and results from the \co\ maps are less reliable in the shaded region. On average, the Virgo galaxies are more \h2-rich at all radii, along with a steeper radial gradient. {\bf Right:} The radial profiles of star formation rate surface density for the galaxies in our sample, calculated using the geometric mean method and normalized by R$_{25}$. The statistical error bars for each profile are plotted. The individual galaxy profiles are in the background. We see an enhancement in the star formation rates near the centre of Virgo galaxies, similar to the behaviour of the \h2\ distribution. However, there are also hints of a truncation in their disks at large radii.}\label{fig-h2_sfr}
\end{figure*}
\subsection{\h2\ Radial Profiles - Enhancement in Virgo Galaxies}
\par
The \h2\ radial profiles are presented in the left plot of Figure~\ref{fig-h2_sfr}. With the small number of CO detected field galaxies and the short range over which we have enough galaxies to generate an average radial profile, the remaining figures will only show a comparison between the Virgo sample and the combined field and group samples. Comparing between the Virgo and non-Virgo samples, we see that the \h2\ disk is not significantly truncated for Virgo galaxies. In fact, the \h2\ surface density is enhanced near the centre. This is consistent with results from the integrated measurements \citep{Mok16}, where we noted an increase in the \h2\ gas mass for the Virgo galaxies in this \hi\ flux selected sample. 
\par
As discussed in Section~\ref{sec-res_hiprofiles}, the NGLS is a \hi-flux selected survey and likely contains galaxies not yet strongly affected by the cluster environment. As a result, our sample may not be directly comparable to those from the Herschel Virgo Cluster Survey (HeViCS) or the Herschel Reference Survey (HRS), which contain a large population of \hi-deficient objects that are also often \h2-deficient. Our results suggest that for recently infalling galaxies or galaxies in less extreme environments, molecular gas can be enhanced rather than reduced by the cluster environment.
\par
A comparison of the \h2\ and \hi\ profiles for Virgo galaxies suggests that while the environment can play a role in truncating the \hi\ disks in these galaxies, some of the gas may lose angular momentum and move towards the centre, into a region where it is more easily converted into molecular form. Simulations of light to moderate ram pressure stripping with cooling processes have shown that low density gas could be preferentially removed from the outskirts while enhancing the amount of high density gas near the centre \citep{Tonnesen09, Bekki114}. Observations of Hickson Compact Groups have shown hints of an enhancement in the molecular gas content compared to isolated galaxies \citep{Martinez-Badenes12}. Changes in the molecular gas distribution have been observed for the galaxies in the Abell 1367 cluster ($z=0.022$), with some galaxies showing signs of \h2\ enhancements \citep{Scott13, Scott15}.
\par
Galaxy interactions, which are more common in the cluster environment, may also aid in this process. For example, results from the AMIGA sample of isolated galaxies suggests a possible molecular gas enhancement for interacting galaxies of approximately 0.2$-$0.3 dex \citep{Lisenfeld11}. 
\subsection{SFR Radial Profiles}
\par
We present the radial profiles of the \ha\ derived star formation rates in the right plot of Figure~\ref{fig-h2_sfr}. Near the central regions, there is an enhancement in the star formation rate density for the Virgo sample, similar to the trends from the \h2\ profiles. The consistency between these two datasets is likely caused by the strong link between molecular gas and star formation. Previous observations have also shown that interactions can drive increases to the star formation rate \citep{Kennicutt87, Ellison13}, which may also contribute to the increase in the star formation rate for the Virgo sample, as several of our galaxies are in close pairs (NGC4567/NGC4568).
\par
At large radii, there are hints of a possible truncation for Virgo galaxies, which was also seen in \citet{Koopmann06}. The behaviour of the star formation rate profiles in this region (R/R$_{25}>1$) is less reliable, as the measurement comes from places outside where the \ha\ flux is normally observed for these galaxies. The large scatter in the distribution of individual galaxy profiles is likely due to imperfect removal of other sources (such as stars) and potential differences in the neighbourhoods of these galaxies.
\subsection{Environmental Trends in the Spatial Distribution of the ISM}
\par
\begin{figure*}
	\centering
	\includegraphics[width=7.5cm]{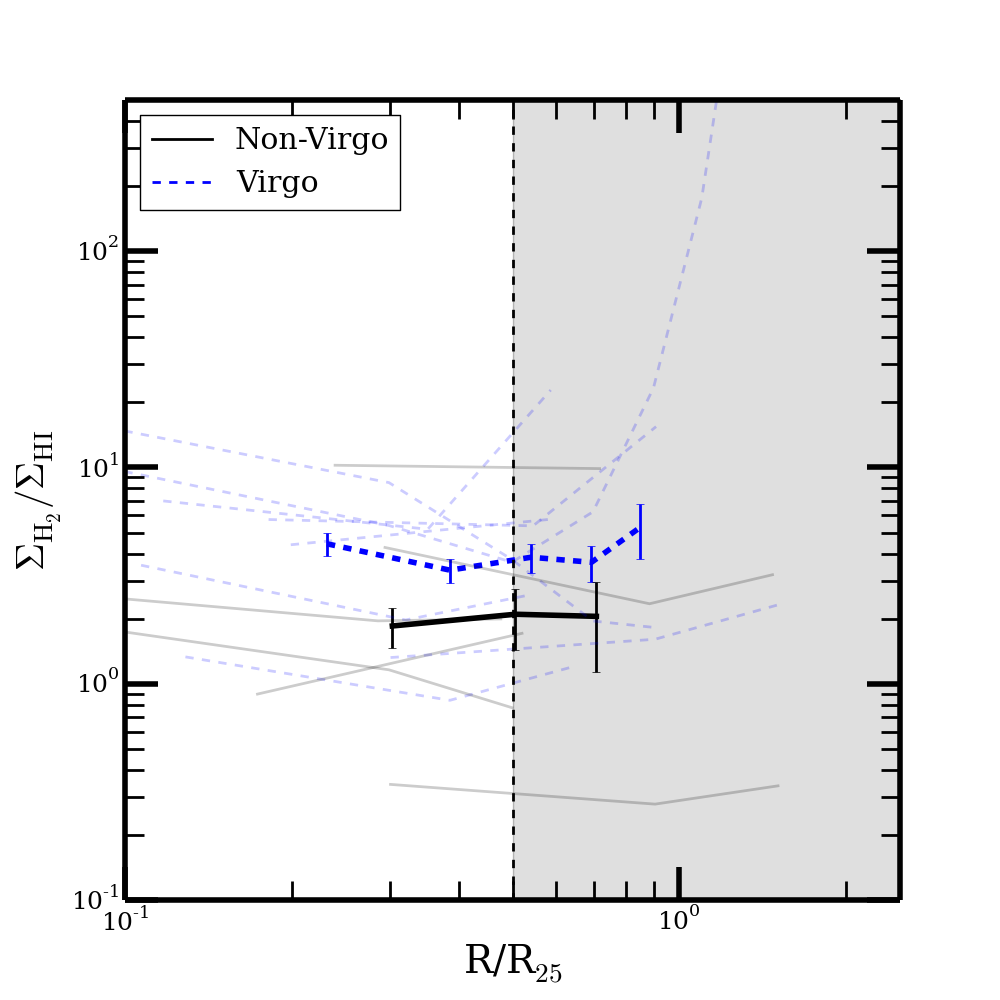}
	\includegraphics[width=7.5cm]{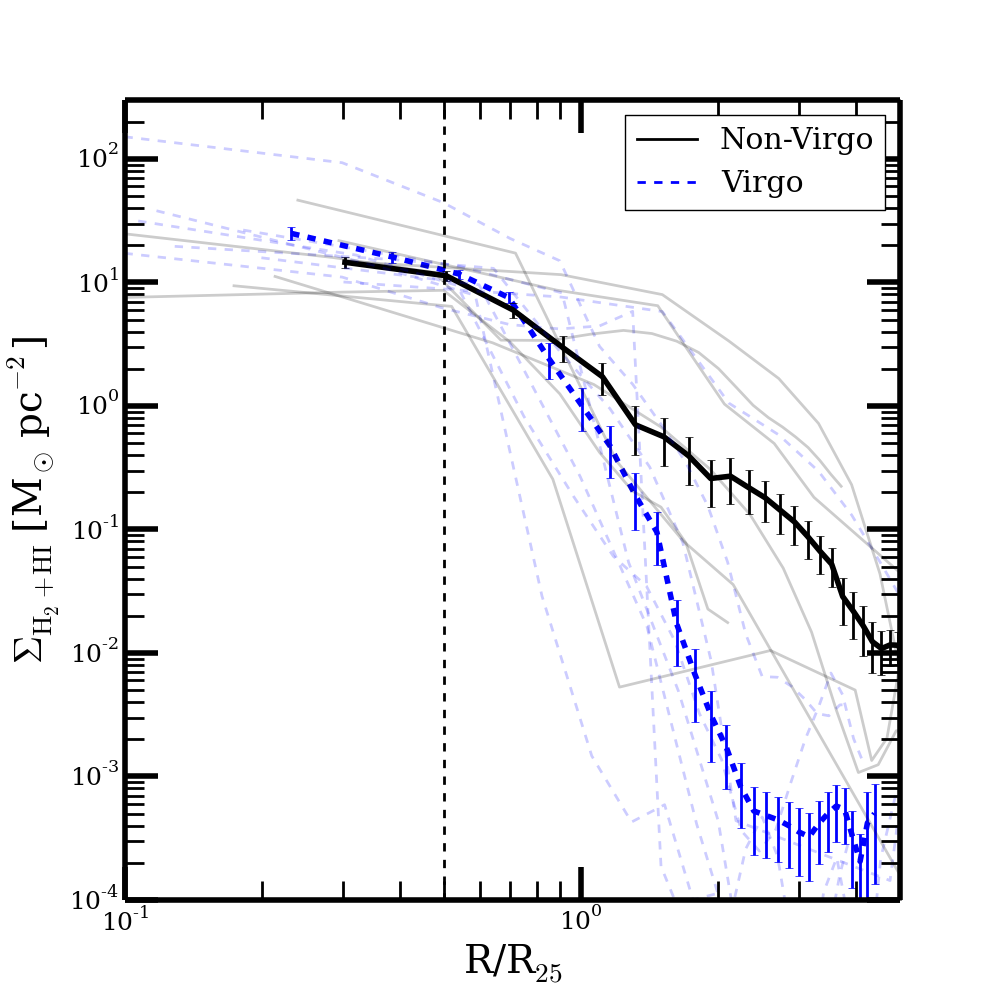}
	\caption{{\bf Left:} The mean radial profiles of the \h2-to-\hi\ ratio for the galaxies in our sample, calculated using the geometric mean method and normalized by R$_{25}$. The statistical error bars for each profile are plotted. The individual galaxy profiles are in the background. The sample is separated into the non-Virgo (black) and Virgo (blue-dashed) populations. The dashed vertical line is at R$_{25}$ = 0.5, the main target area of the NGLS survey, and results from the \co\ maps are less reliable in the shaded region. The \h2-to-\hi ratio shows an enhancement for Virgo galaxies, especially near the centre. {\bf Right:} The radial profiles of the total gas surface density for the galaxies in our sample, calculated using the geometric mean method and normalized by R$_{25}$. The statistical error bars for each profile are plotted. The individual galaxy profiles are in the background. There is a steeper radial distribution for Virgo galaxies compared to non-Virgo galaxies, with more gas concentrated near the centre. Ram pressure stripping is likely playing a role in removing low density gas in the outskirts and enhancing the high density gas near the centre.}\label{fig-derived}
\end{figure*}

In Figure~\ref{fig-derived}, we present the results of the \h2/\hi\ and \h2\ + \hi\ profiles for the smaller sample of 15 galaxies with both sets of measurements. We find that the Virgo galaxies generally have a higher \h2/\hi\ ratio than non-Virgo galaxies, which is also seen in a comparison of the integrated measurements from \citet{Mok16}. The non-Virgo galaxies have a relatively flat profile, at a value of around unity in the region where both \h2\ and \hi\ are detected. Due to the small number of \co\ detected galaxies for the non-Virgo sample and the procedure used in the creation of the radial profiles, we do not probe this ratio in the central region. For the Virgo galaxies, the \h2/\hi ratio rises towards the centre as expected, marking the transition to a molecular gas dominated regime. On the other hand, the slight increase in the \h2/\hi ratio for Virgo galaxies at large radii is likely not significant, as it falls in the region beyond the main NGLS survey area.
\par
For the total gas surface density (\h2\ + \hi) profile, we find differences between the Virgo and non-Virgo samples. Near the centre, the non-Virgo sample have a flatter profile than the Virgo sample, due to the strong contribution from the molecular gas component. In the outskirts, the truncation in the \hi\ disks observed for the profiles of Virgo galaxies is largely responsible for their steep total gas profiles. Steeper \h2\ profiles have also been found for Virgo galaxies from the HeViCS \citep{Pappalardo12}, especially for \hi-deficient galaxies, and fits with the idea that ram pressure stripping may be preferentially removing gas in the outskirts and enhancing the high density gas near the centre, as discussed for the \h2\ surface density profiles.
\subsection{Depressed Star Formation Efficiency at All Radii in Virgo Spirals}
\par
\begin{figure}
	\includegraphics[width=7.5cm]{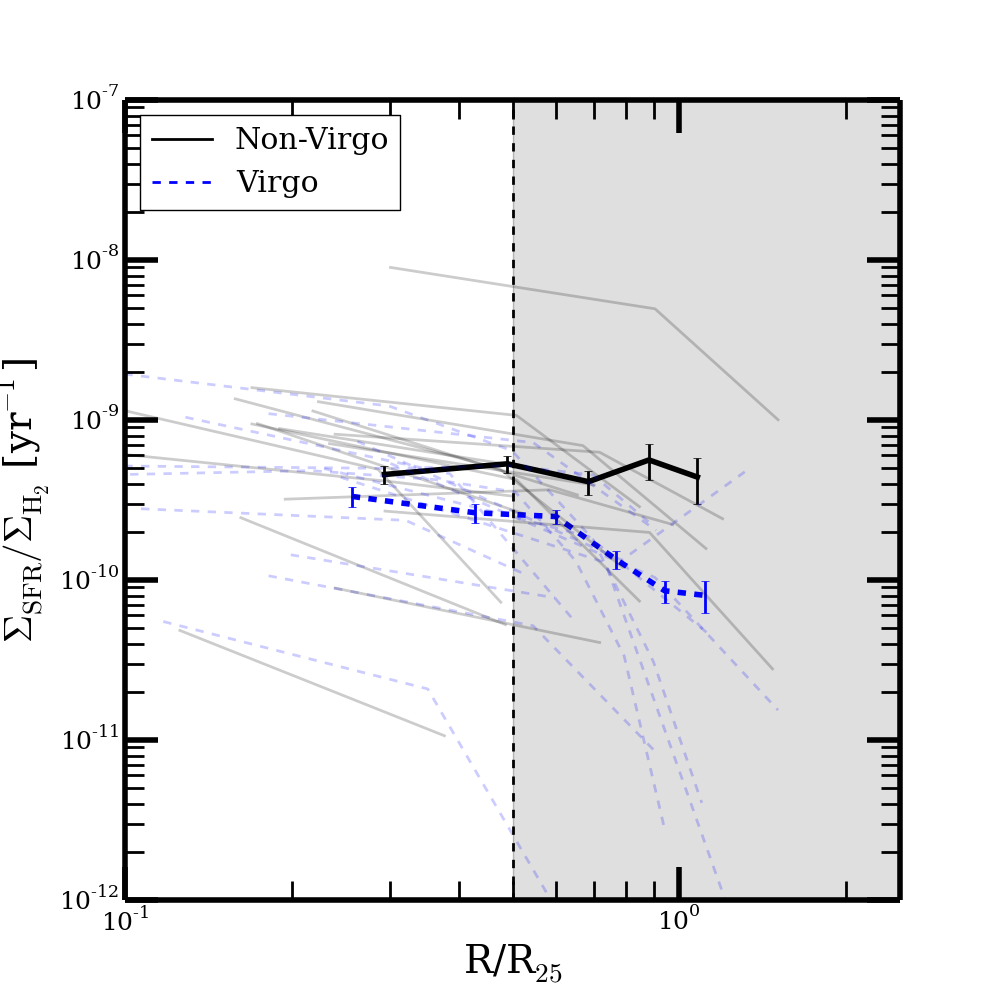}
	\caption{The SFR/\h2\ surface density radial profiles for the galaxies in our sample, calculated using the geometric mean method and normalized by R$_{25}$. The statistical error bars for each profile are plotted. The individual galaxy profiles are in the background. The sample is separated into the non-Virgo (black) and Virgo (blue-dashed) populations. The dashed vertical line is at R$_{25}$ = 0.5, the main target area of the NGLS survey, and results from the \co\ maps are less reliable in the shaded region. Both profiles show a relatively flat trend with radius, with the Virgo galaxies offset at a lower star formation efficiency (or longer molecular gas depletion time), similar to global results presented in \citet{Mok16}.}\label{fig-derived_sfrh2}
\end{figure}

With the \ha-derived star formation rate and the \h2\ data, we can calculate the star formation efficiency (or its reciprocal, the molecular gas depletion time). The results are presented in Figure~\ref{fig-derived_sfrh2}. From the SFR/\h2\ plot, we find that Virgo galaxies generally have a lower star formation efficiency. This result has also been seen in the integrated measurements from \citet{Mok16}. The profile is relatively flat in the inner regions for both samples, with a decrease at large radii for the Virgo sample. This decline in the star formation efficiency may not be physically significant, as it sits outside of the main region of the HARP instrument in the NGLS survey and is in the regime where noise corrections to the \co\ data become important.
\par
The flatness in the radial profiles of the star formation efficiency has been seen in previous studies, which have found near constant molecular gas depletion times for spiral galaxies \citep{Leroy08, Bigiel11, Leroy12}. The absolute values are also in line with these previous results, which have usually found gas depletion times in the $\sim$2 Gyr range. As stated previously, the reduction in the star formation efficiency at larger radii is likely due to the limitations in the survey sensitivity and mapping area.
\par
We also find an average offset in the SFR/\h2\ plot between the Virgo and non-Virgo profiles of $\sim40\%$, which was also present in the integrated measurements from \citet{Mok16}. While Virgo galaxies generally have higher molecular gas masses than non-Virgo galaxies, they are relatively inefficient at turning this fuel into stars. One possibility to explain this result is that there are other factors in the cluster environment, such as difference in the molecular gas properties (pressure, temperature) or larger-scale dynamical stabilization \citep{Martig09} that may be inhibiting the process of star formation in our Virgo sample. A related effect is the possible trend of star formation efficiency with stellar masses \citep{Saintonge11b, Schruba11}, where higher mass galaxies have lower star formation efficiencies. Our Virgo sample has a slightly higher mean stellar mass (log M$_* = 10.05\pm0.12$) compared to the non-Virgo sample (log M$_* = 9.87\pm0.07$), which may contribute to this observed difference. However, given the large scatter in these relationships and the low mean stellar mass of our sample, it is hard to quantify its effect on the star formation efficiency. A final reason may be metallicity effects \citep{Schruba11}, which could affect the observed CO intensity and hence the SFR-to-CO ratio. In this paper, as in other papers in the series, we have used a constant CO-to-\h2\ conversion factor and a constant line ratio between \co\ and CO $J=1-0$. Since most of our galaxies are normal spiral galaxies in a small range of stellar masses, it seems unlikely they would have metallicities of more than a factor of 2 below solar, where these effects become significant \citep{Wilson95, Bolatto08}.
%
%
\section{Conclusions}
\par
We present an analysis of the radial profiles of a sample of 43 \hi-flux selected spiral galaxies from the JCMT Nearby Galaxies Legacy Survey (NGLS) with resolved JCMT \h2\ and/or VLA \hi\ maps. This sample includes field, group, and Virgo Cluster galaxies, and their gaseous disks can act as important probes for environmental effects. Using this gas-rich sample of spiral galaxies, we find significant environmental variations in the spatial distribution of the ISM, including the \hi\ and \h2\ radial profiles, as well as in their star formation properties.
\begin{itemize}
	\item We confirm that the \hi\ disks are truncated in the Virgo sample, compared to the non-Virgo sample. This result is well-known from previous studies of the atomic gas distribution inside clusters \citep{Cayatte94}, but we also observe this same effect for these relatively gas-rich galaxies. Comparing between the smaller group and field samples, we find that there is a decrease in the \hi\ disk size for group galaxies compared to field galaxies, which may be related to the trends in \hi\ properties with group masses \citep{Jaffe16} and suggests that the environment affects the \hi\ gas even in moderate density regions.
	\item We find that the \h2\ distribution is enhanced for Virgo galaxies, especially near the centre, which results in a significantly higher \h2\ to \hi\ ratio for Virgo galaxies. This matches the results found using integrated measurements of these galaxies in \citet{Mok16}. The steeper radial profile for the total gas content (\hi\ + \h2), even for these gas-rich objects, suggests that the environment may play a role in determining the spatial distribution of the ISM inside these galaxies. The most likely scenario is moderate ram pressure stripping that preferentially removes low-density gas in the outskirts of these spiral galaxies and aids in the creation of higher density gas near the centre \citep{Tonnesen09}. Another possibility is interactions in the Virgo sample, as enhancements in molecular gas have also been seen in interacting isolated galaxies \citep{Lisenfeld11} and in compact group galaxies \citep{Martinez-Badenes12}.
	\item The \ha\ star formation rate data for our galaxies show similar trends as the \h2\ distribution. There is an enhancement in the star formation rate surface density for Virgo galaxies near the centre, which is likely related to the increase in the molecular gas surface density. Interactions in the Virgo sample, as discussed for the \h2\ profiles, may also enhance star formation \citep{Kennicutt87, Ellison13}. We also find hints of a \ha\ truncation at large radii for Virgo galaxies, which has been found in other surveys of cluster spirals \citep{Koopmann06}.
	\item We find that the star formation efficiency (SFR/\h2) is relatively constant with radius for the Virgo and non-Virgo samples. This is consistent with the near constant molecular gas depletion times found in other surveys for normal spiral galaxies \citep{Bigiel11, Leroy12}. However, the star formation efficiency for the Virgo galaxies is on average $\sim40\%$ lower than non-Virgo galaxies. This variation may be the result of differences in the star formation process for Virgo galaxies or variations in the molecular gas properties (pressure, temperature, metallicity).
\end{itemize}

%
%
\section{Acknowledgments}
\par
The research of CDW is supported by grants from NSERC (Canada). JHK acknowledges financial support from the Spanish Ministry of Economy and Competitiveness (MINECO) under grant number AYA2013-41243-P.
\par
The James Clerk Maxwell Telescope is operated by the East Asian Observatory on behalf of The National Astronomical Observatory of Japan, Academia Sinica Institute of Astronomy and Astrophysics, the Korea Astronomy and Space Science Institute, the National Astronomical Observatories of China and the Chinese Academy of Sciences (Grant No. XDB09000000), with additional funding support from the Science and Technology Facilities Council of the United Kingdom and participating universities in the United Kingdom and Canada. The National Radio Astronomy Observatory is a facility of the National Science Foundation operated under cooperative agreement by Associated Universities, Inc. This work is based [in part] on archival data obtained with the Spitzer Space Telescope, which is operated by the Jet Propulsion Laboratory, California Institute of Technology under a contract with NASA. Support for this work was provided by an award issued by JPL/Caltech. We acknowledge the usage of the HyperLeda database (http://leda.univ-lyon1.fr). This research has made use of the GOLDMine Database. This research has made use of data from HRS project. HRS is a Herschel Key Programme utilising Guaranteed Time from the SPIRE instrument team, ESAC scientists and a mission scientist. The HRS data was accessed through the Herschel Database in Marseille (HeDaM - http://hedam.lam.fr) operated by CeSAM and hosted by the Laboratoire d'Astrophysique de Marseille.
\par
\par
The Digitized Sky Surveys were produced at the Space Telescope Science Institute under U.S. Government grant NAG W-2166. The images of these surveys are based on photographic data obtained using the Oschin Schmidt Telescope on Palomar Mountain and the UK Schmidt Telescope. The plates were processed into the present compressed digital form with the permission of these institutions. The National Geographic Society - Palomar Observatory Sky Atlas (POSS-I) was made by the California Institute of Technology with grants from the National Geographic Society. The Second Palomar Observatory Sky Survey (POSS-II) was made by the California Institute of Technology with funds from the National Science Foundation, the National Geographic Society, the Sloan Foundation, the Samuel Oschin Foundation, and the Eastman Kodak Corporation. The Oschin Schmidt Telescope is operated by the California Institute of Technology and Palomar Observatory.The UK Schmidt Telescope was operated by the Royal Observatory Edinburgh, with funding from the UK Science and Engineering Research Council (later the UK Particle Physics and Astronomy Research Council), until 1988 June, and thereafter by the Anglo-Australian Observatory. The blue plates of the southern Sky Atlas and its Equatorial Extension (together known as the SERC-J), as well as the Equatorial Red (ER), and the Second Epoch [red] Survey (SES) were all taken with the UK Schmidt.
%
%
\bibliographystyle{mn2e}
\bibliography{./master}
\clearpage
%
\end{document}